\documentclass[11pt]{article}
\pdfoutput=1 
\usepackage{./aux/jheppub}

\usepackage[T1]{fontenc}

\usepackage{graphicx}
\usepackage{amsfonts,amsmath,amssymb}
\usepackage{verbatim}
\usepackage{array}
\usepackage{mathrsfs}
\usepackage{mathrsfs}
\usepackage{yfonts}
\usepackage{dsfont}
\usepackage{bbm}
\usepackage{colonequals}
\usepackage{amscd}

\usepackage{subcaption}

\usepackage{wrapfig}

\usepackage{suffix,mathtools,cancel,bbm}

\usepackage{multirow}

\usepackage{enumerate}

	\usepackage{tikz, tikz-3dplot, pgfplots}
	\usepackage{tkz-graph}
	\usetikzlibrary[positioning,patterns] 
	\usepackage{genyoungtabtikz}

\newcommand{\bn}{\begin{enumerate}}
\newcommand{\en}{\end{enumerate}}

\def\CN{{\cal N}}
\def\CO{{\cal O}}

\def\CS{{\cal S}}

\newcommand{\beq}{\begin{equation}}
\newcommand{\eeq}{\end{equation}}

\newcommand\nn{\nonumber}

\newcommand{\cA}{\mathcal{A}}
\newcommand{\cB}{\mathcal{B}}
\newcommand{\cC}{\mathcal{C}}

\newcommand{\cE}{\mathcal{E}}
\newcommand{\cF}{\mathcal{F}}

\newcommand{\cI}{\mathcal{I}}

\newcommand{\cK}{\mathcal{K}}

\newcommand{\cM}{\mathcal{M}}
\newcommand{\cN}{\mathcal{N}}
\newcommand{\cO}{\mathcal{O}}

\newcommand{\cR}{\mathcal{R}}
\newcommand{\cS}{\mathcal{S}}

\numberwithin{equation}{section}

\def\bea{\begin{eqnarray}}
\def\eea{\end{eqnarray}}

\DeclarePairedDelimiterX\MeijerM[3]{\lparen}{\rparen}%
{\begin{smallmatrix}#1 \\ #2\end{smallmatrix}\delimsize\vert\,#3}

\newcommand\MeijerG[8][]{%
  G^{\,#2,#3}_{#4,#5}\MeijerM[#1]{#6}{#7}{#8}}

\WithSuffix\newcommand\MeijerG*[7]{%
  G^{\,#1,#2}_{#3,#4}\MeijerM*{#5}{#6}{#7}}

\def\tr{\mathop{\mathrm{tr}}\nolimits}

\def\cN{\mathcal{N}}

\def \beg#1{\begin{#1}} 

\def \bea{\beg{eqnarray}}
\def \eea{\end{eqnarray}}
\def \ee{\end{equation}}

\def \restr#1#2{{\left.\kern-\nulldelimiterspace#1\vphantom{\big|}\right|_{#2}}}

\def \nn{\nonumber}

\newcommand{\ba}[1]{\begin{align} #1 \end{align} }

\newcommand{\bs}[1]{\begin{split} #1 \end{split} }

\title{\boldmath Anomaly Inflow for M5-branes on Punctured Riemann Surfaces}

\author[a]{Ibrahima Bah,}
\author[a]{Federico Bonetti,}
\author[b]{Ruben Minasian,}
\author[c]{and Emily Nardoni} 
\affiliation[a]{Department of Physics and Astronomy, Johns Hopkins University, 3400 North Charles Street, Baltimore, MD 21218, USA}
\affiliation[b]{Institut de Physique Th\'{e}orique, Universit\'{e} Paris Saclay, CNRS, CEA, F-91191, Gif-sur-Yvette, France}
\affiliation[c]{Mani L. Bhaumik Institute for Theoretical Physics, Department of Physics and Astronomy, University of California, Los Angeles,  CA 90095, USA}

\emailAdd{iboubah@jhu.edu, fbonett3@jhu.edu, ruben.minasian@ipht.fr, enardoni@ucla.edu}

\abstract
{
We derive the anomaly polynomials of 4d $\CN=2$ theories that are obtained by wrapping M5-branes on a Riemann surface with arbitrary regular punctures, using anomaly inflow in the corresponding M-theory setup. Our results match the known anomaly polynomials for the 4d $\CN=2$ class $\CS$ SCFTs. In our approach, the contributions to the 't Hooft anomalies due to boundary conditions at the punctures are determined entirely by $G_4$-flux in the 11d geometry. This computation provides a top-down derivation of these contributions that utilizes the geometric definition of the field theories, complementing the previous field-theoretic arguments. 
}

\usepackage{etoolbox}
\appto\appendix{\addtocontents{toc}{\protect\setcounter{tocdepth}{1}}}

\appto\listoffigures{\addtocontents{lof}{\protect\setcounter{tocdepth}{1}}}
\appto\listoftables{\addtocontents{lot}{\protect\setcounter{tocdepth}{1}}}

\begin{document} 

\maketitle
\flushbottom



\section{Introduction}

Geometric engineering has become a standard tool for constructing and exploring 
quantum field theories, especially in their strong coupling regimes.  
A  large class of generically strongly coupled QFTs
in four dimensions is realized in M-theory
by wrapping a stack of M5-branes on a Riemann surface with defects.
These constructions fit in the larger framework of the class $\cS$ program,
in which 4d QFTs are obtained by dimensional reduction
of a 6d SCFT, generically with a partial topological
twist.
In this work we focus on the case of the 6d (2,0) theory of type
$A_{N-1}$, which is the worldvolume theory on a stack of $N$ M5-branes.
Depending on the choice of 
 twist, the theories of class $\CS$ can preserve $\CN=2$ or $\CN=1$ supersymmetry\footnote{4d $\CN=4$ is obtained by compactifying on a torus with no twist.}. The $\CN=2$ theories  were first constructed in \cite{Gaiotto:2009we,Gaiotto:2009hg}, building on work in \cite{Witten:1997sc}. A large class of $\CN=1$ theories of class $\CS$ were constructed in \cite{Bah:2011vv,Bah:2012dg}, building on work in \cite{Maruyoshi:2009uk,Benini:2009mz,Bah:2011je}.
Strong evidence for the existence of these SCFTs is  the construction of their large-$N$ gravity duals. The holographic duals of the $\CN=2$ theories were identified in \cite{Gaiotto:2009gz}, and for the $\CN=1$ theories in \cite{Bah:2011vv,Bah:2012dg,
Bah:2013qya,
Bah:2015fwa}.

't Hooft anomalies provide crucial insight into the properties of QFTs, and are especially useful observables in the study of strongly coupled theories\footnote{Throughout, we refer to anomalies in background (rather than dynamical) gauge or gravity fields as 't Hooft anomalies.}.  In an interacting SCFT, anomalies are related to central charges by the superconformal algebra \cite{Kuzenko:1999pi,Anselmi:1997am};
in a free theory,  they directly specify the matter content. Thus, they provide a measure of the degrees of freedom in a QFT.   The anomalies of a $d$-dimensional QFT can be organized in a $(d+2)$-form  known as the anomaly polynomial, which is a polynomial in the curvatures of background gauge and gravitational
fields associated to global symmetries \cite{AlvarezGaume:1983ig,AlvarezGaume:1984dr,Bardeen:1984pm}.   The geometric nature 
of anomalies
makes them especially amenable to computation in geometrically engineered constructions.

The 6-form 't Hooft anomaly polynomial for a 4d theory of class $\cS$
depends on the parent 6d theory,
on the genus-$g$, $n$-punctured  Riemann surface $\Sigma_{g,n}$
used in the compactification,
and on the boundary conditions for the 6d theory
at the punctures.
The total anomaly polynomial
$I_6^{\rm CFT}$ can be decomposed 
as a  sum of a ``universal'' or ``bulk'' term,
and of individual terms for each puncture \cite{Bah:2018gwc},
	\ba{
	I_6^{\text{CFT}} = I_6^{\text{CFT}}(\Sigma_{g,n} ) + \sum_{\alpha = 1}^n I_6^{\text{CFT}}(P_\alpha)\ .
	}
The bulk term $I_6^{\text{CFT}}(\Sigma_{g,n} )$ depends 
on the surface only through its Euler characteristic, 
$\chi (\Sigma_{g,n})= - 2(g-1)-n$,
and is insensitive to the choice of boundary
conditions at the punctures.
This contribution for the $\CN=2$ theories of class $\CS$ was first computed in \cite{Gaiotto:2009gz} using S-duality, and can be computed by integrating
the 8-form anomaly polynomial of the 6d theory over the Riemann surface
\cite{Alday:2009qq,Bah:2011vv,Benini:2009mz,Bah:2012dg}.

The individual puncture contribution
$I_6^{\text{CFT}}(P_\alpha)$ depends on the choice of boundary
conditions at the puncture $P_\alpha$,
and contains information about the 't Hooft anomalies of the
flavor
symmetry associated to it. 
These contributions can be obtained by S-duality and anomaly matching arguments \cite{Gaiotto:2009gz,Chacaltana:2010ks,Chacaltana:2012zy,Tachikawa:2015bga}.

The main goal of this paper is a first-principles  derivation of the anomalies of the
$\CN=2$ class
$\CS$ theories  of type $A_{N-1}$ from their geometric construction
via M5-branes. Using anomaly inflow in M-theory,
we determine both the
bulk term $ I_6^{\text{CFT}}(\Sigma_{g,n} )$
and the puncture term $I_6^{\text{CFT}}(P_\alpha)$,
for any regular puncture.
Our analysis is inspired and motivated by the holographic
duals of these theories \cite{Gaiotto:2009gz}.
The present work is a follow up to \cite{Bah:2018jrv}, where the results of the computation and main features of the derivation were presented.

The outline of the rest of the paper  is as follows.
In section \ref{section_strategy} we provide an overview of the main
strategy used in the computation of the inflow anomaly polynomial.
In section \ref{sec_setup} we describe in greater detail the
M5-brane setup, and we discuss the bulk
contribution to   anomaly inflow.
Section \ref{sec_punctures} is devoted to the discussion of the local
geometry and $G_4$-flux configuration near a puncture.
These data are used in section \ref{sec_puncture_anomaly}
to compute the puncture contribution to anomaly inflow.
In section \ref{sec_comparison_with_CFT} we compare the total inflow
result with the known CFT anomaly polynomial.
In the conclusion we summarize our findings and 
discuss future directions.
Some technical aspects of our derivation
are relegated to the appendices, together with 
 useful background material.



\section{Outline of Computation}
\label{section_strategy}

Our goal is an anomaly-inflow derivation of the 't Hooft anomaly polynomial
of 4d $\cN = 2$ class $\cS$ theories with regular punctures.
In this section we provide a summary and overview of the strategy 
used in the main computations in this paper.

Anomaly cancellation for M5-branes in M-theory
was analyzed in \cite{Duff:1995wd,Witten:1996hc,Freed:1998tg,Harvey:1998bx,Intriligator:2000eq,Yi:2001bz}. The quantum anomaly
generated by the chiral degrees of freedom localized on the M5-brane stack
is cancelled by a classical inflow from the 11d ambient space.
In section \ref{sec_classI12}, we briefly review
this mechanism and argue that it can be neatly
summarized by introducing a 12-form characteristic class $\cI_{12}$.
The class $\cI_{12}$ is related via standard descent relations
to the classical anomalous variation of the 11d action,
see \eqref{deltaSM}, \eqref{I12_descent} below.
Upon integrating $\cI_{12}$ along the $S^4$ surrounding the M5-brane
stack,
one recovers the 8-form anomaly polynomial
of the 6d $(2,0)$ theory of type $A_{N-1}$,
up to the decoupling of center-of-mass modes.

In this work we study 4d theories obtained by considering
an M5-brane stack with worldvolume $W_6= W_{4} \times \Sigma_{g,n}$,
where $W_{4}$ is external 4d spacetime
and $\Sigma_{g,n}$ is a Riemann surface of genus $g$ with $n$ punctures.
In section \ref{sec_no_punctures},
we   consider the case without punctures, and argue that
the 6-form anomaly polynomial
of the resulting 4d theory can be computed by integrating
$\cI_{12}$ on a suitable 6d space $M_6$,
which is an $S^4$ fibration over $\Sigma_{g,0}$.
In section \ref{sec_inclusion_of_punctures},
we outline a two-step procedure for introducing punctures.
Firstly, one constructs a modified version of $M_6$,
by excising $n$ small disks from the Riemann surface,
together with the $S^4$ fibers on top of them.
Secondly, the ``holes'' in $M_6$ are ``filled''
with new geometries supported by non-trivial $G_4$-flux.
The latter encode all data about the punctures.

\subsection{Anomaly Inflow and the Class $\cI_{12}$}
\label{sec_classI12}
Consider a stack of $N$ coincident M5-branes with 
a smooth 6d worldvolume $W_6$.
The 11d tangent bundle of the ambient space $M_{11}$,
restricted to  $W_6$, 
decomposes as 
\ba{
\label{TM11_decomposition}
TM_{11} |_{W_6}= TW_{6} \oplus NW_6 \ ,
}
where $TW_{6}$, $NW_6$ are the tangent bundle and normal bundle
to the M5-brane stack, respectively.
The normal bundle $NW_6$ is isomorphic to a small
tubular neighborhood of $W_6$ inside $M_{11}$.
From this point of view, the M5-brane stack sits at the origin
of the $\mathbb R^5$ fibers of $NW_6$,
which encode the five directions transverse to the stack.  The normal bundle admits an $SO(5)$ structure group.  It induces an 
$SO(5)$ action onto the degrees of freedom on the brane; this is identified with the R-symmetry of the quantum field theory living on the branes.

The M5-brane stack acts as a singular magnetic source for the M-theory
4-form flux $G_4$. The Bianchi identity $dG_4 = 0$ is modified to
\beq
\label{dG4_delta}
dG_4 = 2\pi \, N \, \delta^5(y) \, dy^1 \wedge dy^2 \wedge dy^3 \wedge dy^4
\wedge dy^5 \ , 
\eeq
where $y^A$, $A = 1, \dots, 5$, are local Cartesian coordinates
in the $\mathbb R^5$ fibers of $NW_6$, 
and $\delta^5(y)$ is the standard 5d delta function.
The relation \eqref{dG4_delta} should only be considered
as a schematic expression. As explained in \cite{Freed:1998tg,Harvey:1998bx}, \eqref{dG4_delta}  must be improved 
in two respects in order to implement anomaly inflow.

In the first step, we regularize the delta-function singularity
in \eqref{dG4_delta}. This is achieved by 
excising a small tubular neighborhood $B_\epsilon$ of radius $\epsilon$ 
of the M5-brane stack. Next, we introduce a radial bump
function $f(r)$, with $r$ denoting the 
 radial coordinate $r^2 =\delta_{AB} y^A y^B$.
 The function $f$ is equal to $-1$ at $r = \epsilon$,
and approaches $0$ monotonically as we increase $r$.
The relation \eqref{dG4_delta}  is thus replaced by
\beq
d G_4 = 2\pi \, N \, df \wedge {\rm vol}_{S^4} \ .
\eeq
The 4-form ${\rm vol}_{S^4}$ is the volume
form on the $S^4$ surrounding the origin of the $\mathbb R^5$
transverse directions,
normalized   to integrate to $1$.

The second step is to gauge the $SO(5)$ action of the normal bundle. 
This requires that we replace $N \,{\rm vol}_{S^4}$
with a multiple of the global angular form,
\beq \label{dG4_generic}
\frac{dG_4}{2\pi} =  df \wedge E_4 \ .
\eeq
Let us stress that, in our notation, we absorb the factor $N$ inside
$E_4$,
\beq \label{new_E4_normalization}
\int_{S^4} E_4 = N \ .
\eeq
The closed and $SO(5)$ invariant 4-form $E_4$ is constructed
with the coordinates $y^A$ and the 
$SO(5)$ connection
$\Theta_{[AB]}$ on $NW_6$. We refer the reader to appendix \ref{appendix_global_ang_forms}
for the explicit expression of $E_4$.

After excising a small tubular neighborhood $B_\epsilon$ of the M5-brane
stack, the 11d spacetime $M_{11}$ acquires a non-trivial boundary
$M_{10}$ at $r = \epsilon$, which is an $S^4$ fibration over the 
worldvolume $W_6$,
\beq \label{X10boundary}
M_{10} \equiv \partial(M_{11} \setminus B_\epsilon)  \ , \qquad
S^4 \hookrightarrow M_{10} \rightarrow W_6 \ .
\eeq
The M-theory effective action $S_{\rm M}$ on $M_{11} \setminus B_\epsilon$
is no longer invariant under diffeomorphisms and gauge transformations of the M-theory
3-form $C_3$. The classical variation of the action $S_M$
under such a transformation takes the form
\beq \label{deltaSM}
\frac{\delta S_{\rm M}}{2\pi} = \int_{M_{10}} \cI_{10}^{(1)} \ ,
\eeq
where $\cI_{10}^{(1)}$ is a 10-form proportional to the
gauge parameters. By virtue of the Wess-Zumino
consistency conditions, the quantity 
$\cI_{10}^{(1)}$ is related via descent to a formal
12-form characteristic class,
\beq \label{I12_descent}
d\cI_{10}^{(1)} = \delta \cI_{11}^{(0)} \ , \qquad
d\cI_{11}^{(0)} =  \cI_{12} \ .
\eeq
We are adopting a standard descent notation,
with the superscript $(0)$, $(1)$ indicating the power of the
variation parameter.
The class $\cI_{12}$
originates from the topological couplings in the M-theory
effective action, and is given by
\beq \label{I12_def}
\mathcal I_{12}  =    - \frac 16 \, E_4 \wedge E_4 \wedge E_4
- E_4 \wedge X_8  \ .
\eeq
We refer the reader to
appendix \ref{sec_I12_derivation} for a review of the derivation,
based on 
\cite{Freed:1998tg,Harvey:1998bx}.
In \eqref{I12_def}, $X_8$ is the 8-form
characteristic class
\beq \label{I8def}
X_8 = \frac{1}{192} \bigg[ p_1(TM_{11}) ^2 - 4 \, p_2(TM_{11}) \bigg] \ ,
\eeq
where $TM_{11}$ is the tangent bundle
to 11d spacetime $M_{11}$, and $p_i (TM_{11})$ denote its 
Pontryagin classes. Let us stress that
a pullback to $M_{10}$ is implicit in \eqref{I12_def}.

The relevance of the 12-form characteristic class $\cI_{12}$
stems from the fact that, upon integrating it along the $S^4$
transverse to the M5-brane stack,
we obtain the inflow
anomaly polynomial of the 6d theory living on the stack \cite{Freed:1998tg,Harvey:1998bx},
\beq \label{I8inflow}
I_8^{\rm inflow} = \int_{S^4} \cI_{12} \ .
\eeq
Notice that \eqref{I8inflow} makes use implicitly
of the fact that descent and integration over $S^4$ commute.
We offer an argument for the previous statement in appendix \ref{sec_descent_commutes}.

The anomaly polynomial $I_8^{\rm inflow}$ cancels
against the quantum anomalies of the 
chiral degrees of freedom on the M5-brane stack.
In the IR, the latter are organized into the
interacting degrees of freedom of the 6d $(2,0)$ theory of type
$A_{N-1}$, together with one free 6d $(2,0)$ tensor multiplet,
related to the center of mass of the M5-brane stack.
We may then write
\beq
I_8^{\rm inflow} + I_8^{\rm CFT} + I_8^{\rm decoup} = 0 \ ,
\eeq
where $I_8^{\rm CFT}$ is the anomaly polynomial
of the interacting $(2,0)$ theory,
and $I_8^{\rm decoup}$ is the anomaly polynomial
of a free   $(2,0)$ tensor multiplet.

\subsection{Four-Dimensional Anomalies from Integrals of $\cI_{12}$}
\label{sec_no_punctures}
The discussion of the previous subsection is readily
specialized to the case in which the M5-brane worldvolume
is $W_6 = W_4 \times \Sigma_{g,0}$, where $W_4$ is external 4d spacetime,
and $\Sigma_{g,0}$ is a Riemann surface of genus $g$ without punctures.
In such a setup, the structure group of the 
normal bundle $NW_6$ is reduced from 
$SO(5)$ to $SO(2) \times SO(3)$ or $SO(2) \times SO(2)$,
for compactifications preserving 4d $\cN = 2$ or $\cN =1$
supersymmetry, respectively.
A more detailed explanation of this point
is found in section \ref{normal} below. 

The space $M_{10}$ introduced in \eqref{X10boundary} is now
an $S^4$ fibration over $W_4 \times \Sigma_{g,0}$.
The connection splits into an external part with legs on $W_4$
and an internal part with legs on $\Sigma_{g,0}$.
The external part of the connection
on $NW_6$ is  a background
gauge field for the continuous global symmetries of the 4d
field theory. When these background gauge
fields are turned off, the space $M_{10}$ decomposes as
the product of $W_4$ and a 6d space, denoted $M_6^{n=0}$
to emphasize that we are considering a setup with no punctures.
The space $M_6^{n=0}$ is an $S^4$ fibration over $\Sigma_{g,0}$,
\beq \label{M6_no_punct}
S^4 \hookrightarrow M_{6}^{n=0} \rightarrow \Sigma_{g,0} \ .
\eeq
It
is fixed by the supersymmetry conditions
of M-theory, as discussed 
 in section \ref{normal}.
We can now regard $M_{10}$
as an $M_6^{n=0}$ fibration over $W_4$,
\beq \label{M10_fibration}
M_6^{n=0} \hookrightarrow M_{10} \rightarrow W_4 \ .
\eeq
The topology of the above fibration encodes the information
originally contained in \eqref{X10boundary}.

We argue that the inflow anomaly polynomial
$I_{6}^{\rm inflow}$ for the 4d
field theory is given by
\beq \label{anomaly_integral}
I_{6}^{\rm inflow} = \int_{M_{6}^{n=0}} \cI_{12} \ ,
\eeq
with $\cI_{12}$ given in \eqref{I12_def}.
We should bear in mind that,
in analogy with the uncompactified case,
the inflow anomaly polynomial $I_{6}^{\rm inflow}$ 
balances against the contributions of an interacting CFT as well as
of decoupling modes,
\beq \label{CFT_and_decoup}
I_6^{\rm inflow} + I_6^{\rm CFT} + I_6^{\rm decoup} = 0 \ .
\eeq
The decoupling modes are precisely 
those arising from the compactification of a free 6d $(2,0)$
tensor multiplet on $\Sigma_{g,0}$.
We   stress that 
\eqref{CFT_and_decoup} generically fails 
in the case of emergent symmetries in the IR,
in which case $I_6^{\rm inflow}$ might not capture
all the anomalies of the CFT.

\subsection{Inclusion of Punctures}
\label{sec_inclusion_of_punctures}

Let us now outline a general strategy for extending
\eqref{anomaly_integral} to the case of a compactification
of an M5-brane stack 
on
a Riemann surface $\Sigma_{g,n}$ of genus $g$ with $n$ punctures. 
Let $P_\alpha$ be the point on the Riemann surface
where the $\alpha^{\rm th}$ puncture is located,
for $\alpha = 1, \dots,n$.

Our starting point is the 
space  $M_6^{n=0}$ as in \eqref{M6_no_punct}.
Let $D_\alpha$ denote a small disk on the Riemann surface,
centered around the point $P_\alpha$.
We can present the space $M_6^{n=0}$ as
\beq \label{M6_decomposed}
M_6^{n=0} = M_6^{\rm bulk} \cup \bigcup_{\alpha = 1}^n (D_\alpha \times S^4) \ ,
\eeq
where $M_6^{\rm bulk}$ denotes the space
obtained from 
$M_6^{n=0}$ by excising the small disk $D_\alpha$
around each point $P_\alpha$, together with the $S^4$ fiber on top of it.
It follows that $M_6^{\rm bulk}$ is an $S^4$ fibration over~$\Sigma_{g,n}$,
\beq \label{M6bulk_fibration}
S^4 \hookrightarrow M_6^{\rm bulk } \rightarrow \Sigma_{g,n} \ .
\eeq

To introduce punctures, we replace each term
$D_\alpha \times S^4$ in \eqref{M6_decomposed} with a new 
geometry $X_6^\alpha$ that encodes the puncture data.
We denote the   resulting space as $ M_6$,
\beq   \label{eq_collage}
  M_6 = M_6^{\rm bulk} \cup \bigcup_{\alpha = 1}^n X_6^\alpha \ .
\eeq
Smoothness of $  M_6$ constrains the 
gluing of $X_6^\alpha$ onto $M_6^{\rm bulk}$. 
In analogy with \eqref{M10_fibration}, 
the 10d space $M_{10}$ is an $M_6$ fibration 
over external spacetime $W_4$,  
\beq
  M_6 \hookrightarrow   M_{10} \rightarrow W_4 \ .
\eeq
Each  local geometry $X_6^\alpha$ in \eqref{eq_collage}
is supported by a non-trivial $G_4$-flux configuration,
which is encoded in the class $E_4$ on $M_{10}$.
The geometry $X_6^\alpha$, together with $E_4$ near the puncture,
encodes the details of the puncture at $P_\alpha$.
In contrast with \eqref{M6_no_punct},  the space $X_{6}^\alpha$ is \emph{not}   an $S^4$ fibration over a 2d base space.

The class $E_4$ in $\cI_{12}$ is understood as a globally-defined
object on $M_{10}$. In this work we construct  local expressions for
$E_4$, both in
the bulk of the Riemann surface and near each puncture,
which    are then constrained
by regularity and flux quantization.
These conditions turn out to be enough to 
determine the inflow anomaly polynomial.

The structure of $M_6$ in  \eqref{eq_collage}
implies that
the total inflow anomaly polynomial
can be written as a sum of a \emph{bulk contribution},
associated to $M_6^{\rm bulk}$,
and the individual contributions of punctures,
associated to $X_6^\alpha$,
\beq \label{decomposing_inflow}
I_{6}^{\rm inflow} = \int_{  M_6} \cI_{12}  = I_{6}^{\rm inflow} (\Sigma_{g,n})
+ \sum_{\alpha = 1}^n I_{6}^{\rm inflow} (P_\alpha)  \ , 
\eeq
where one has
\beq \label{decomposing_inflow2}
I_{6}^{\rm inflow}(\Sigma_{g,n}) = \int_{   M_6^{\rm bulk}} \cI_{12} \ , \qquad
I_{6}^{\rm inflow} (P_\alpha) = \int_{X_{6}^\alpha} \cI_{12}  \ .
\eeq

Several comments are in order 
regarding the decomposition \eqref{decomposing_inflow}.
First of all, we stress that one should think of 
$M_{6}^{\rm bulk}$ as a space with boundaries.
Accordingly, one has to assign suitable boundary conditions 
 at the punctures for the 
connection
in the fibration
\eqref{M6bulk_fibration}.
Notice also that \cite{Bah:2018gwc}
\beq
I_6^{\rm inflow}(\Sigma_{g,n}) =  \int_{   M_6^{\rm bulk}} \cI_{12} 
=\int_{\Sigma_{g,n}} I_8^{\rm inflow}  \ ,
\eeq
where the integration over $M_6^{\rm bulk}$
is performed by first integrating along the $S^4$ fibers,
and then integrating on $\Sigma_{g,n}$.
The class 
$I_8^{\rm inflow}$ is given by \eqref{I8inflow}
and captures the anomalies of the
6d (2,0) SCFT that lives on a stack of flat M5-branes.

The local geometry $X_{6}^\alpha$ and its $G_4$-flux configuration are 
constrained by several consistency conditions.
As mentioned earlier, we must be able to glue the local geometry $X_{6}^\alpha$  
smoothly onto the bulk geometry $M_6^{\rm bulk}$.
Moreover, the gluing must 
preserve all the relevant symmetries of the 
problem (including the correct amount of supersymmetry).
Section \ref{sec_punctures} below is   devoted 
to describing all the relevant features of the geometries
 $X_{6}^\alpha$ and associated $E_4$ configurations that describe regular punctures
 for $\cN = 2$ class $\cS$ theories.



\section{M5-brane Setup}  
\label{sec_setup}

This section is devoted to the   description of the
M-theory setup of a stack of $N$ M5-branes wrapping a Riemann surface
$\Sigma_{g,n}$ of genus $g$ with $n$ punctures. 
In particular, we recall the properties of the
normal bundle to the M5-branes in this scenario,
and its role in implementing a partial topological
twist of the parent 6d $(2,0)$ theory on $\Sigma_{g,n}$,
which is essential to preserve supersymmetry in four dimensions.
We then discuss the properties of the class $E_4$
and of the $S^4$ fibration $M_6^{n=0}$, introduced in \eqref{dG4_generic} and \eqref{M6_no_punct}. 
We proceed to analyze $M_6^{\rm bulk}$.
This enables us to compute the bulk contribution
to the inflow anomaly polynomial $\cI_6^{\rm inflow}(\Sigma_{g,n})$
according to \eqref{decomposing_inflow}.

\subsection{Normal Bundle to the M5-brane Stack}\label{normal}

The 11d tangent space restricted to the M5-brane worldvolume
decomposes according to \eqref{TM11_decomposition}.
We are interested in the case in which the worldvolume $W_6$
wraps a Riemann surface
$\Sigma_{g,n}$ of genus $g$ with $n$ punctures.
The tangent space to $W_6$ decomposes
according to
\ba{
\label{TW6_decomposition}
TW_6 = T W_4 \oplus T\Sigma_{g,n} \ .
}
The Chern root of
$T\Sigma_{g,n}$
is denoted $\hat t$
and satisfies
\ba{ \label{hatt_integral_no_punct}
\int_{\Sigma_{g,n}} \hat t  = \chi(\Sigma_{g,n}) =  -2(g-1) -n \ .
}

We consider  setups preserving $\cN = 2$
supersymmetry in four dimensions,
in which 
the structure group of $NW_6$  reduces
from $SO(5)$ to $SO(2)  \times SO(3)$.
Accordingly, 
 $NW_6$   decomposes
in a direct sum, 
\beq
\label{NW6_N2_decomposition}
NW_6 = N_{SO(2)} \oplus N_{SO(3)} \ ,
\eeq
where $N_{SO(2)}$ is a bundle over $W_6$ with fiber $\mathbb R^2$
and structure group $SO(2)$,
and $N_{SO(3)}$ is a bundle with fiber $\mathbb R^3$ and
structure group $SO(3)$.
Let $\hat n$ denote the Chern root of  
$N_{SO(2)}$.
We can write
\beq \label{hatn_eq}
\hat  n = - \hat t + \hat n^{\rm 4d} \ ,
\eeq
where $\hat n^{\rm 4d}$ denotes the part of $\hat n$
depending on external spacetime.
The part of $\hat n$ depending on $\Sigma_{g,n}$ is fixed to be $-\hat t$.
This identification amounts to a topological twist of the parent
6d $(2,0)$ $A_{N-1}$ theory compactified on $\Sigma_{g,n}$,
and is 
necessary to preserve 4d $\cN =2$ supersymmetry \cite{Gaiotto:2009hg}. 
The angular directions in the fibers of $NW_6$
are identified with the $S^4$ fiber in   
\eqref{M6_no_punct}, \eqref{M6bulk_fibration} in the absence of punctures
and in the presence of punctures, respectively.

The decomposition \eqref{NW6_N2_decomposition}
suggests a presentation of the $S^4$  
 as an $S^1\times S^2$ fibration
over an interval with coordinate $\mu \in [0,1]$.
This is readily achieved by the following parametrization of
$y^A$, $A = 1,\dots,5$:
\beq \label{mu_equation}
y^{1,2,3} = r \, \mu \, \hat y^{1,2,3} \ , \qquad
(\hat y^1)^2 + (\hat y^2)^2 + (\hat y^3)^2 = 1  \  , \qquad
y^4 + i \, y^5 =r \,  \sqrt{1-\mu^2} \, e^{i\phi} \ .
\eeq
We use the symbol $S^2_\Omega$ for the 2-sphere
defined by the second relation. The isometries of $S^2_\Omega$
are related to 
 the $SU(2)_R$ R-symmetry of the 4d
theory. We refer the symbol $S^1_\phi$
for the circle parametrized by the angle $\phi$.
Throughout this work,
the angle $\phi$ has periodicity $2\pi$.
The isometry of $S^1_\phi$
corresponds to the $U(1)_r$ R-symmetry in four dimensions.
As apparent from \eqref{mu_equation},
the circle $S^1_\phi$ shrinks for $\mu = 1$,
while the 2-sphere $S^2_\Omega$ shrinks for $\mu = 0$.

 The
gauge-invariant differential for the angle $\phi$ reads
\beq \label{Dphi_def}
D\phi = d\phi - \cA  \ ,
\eeq
where $\cA$ is the total connection for the bundle
$N_{SO(2)}$. The field strength of $\cA$ is $\cF = d\cA$,
and $\cF/(2\pi)$ is identified with the Chern
character $\hat n$.
Both $\cA$ and $\cF$ can be split into an internal part,
with legs on the Riemann surface,
and a part with legs along external spacetime.
We use the notation
\beq \label{NSO2_connection}
\cA = A_\Sigma + A_\phi \ , \qquad 
\cF = F_\Sigma + F_\phi \ ,
\eeq
where the first term is the internal piece,
and the second is the external piece.
Thanks to \eqref{hatn_eq},
we have
\beq \label{cF_on_Sigma}
\int_{\Sigma_{g,n}} \frac{\cF}{2\pi} =\int_{\Sigma_{g,n}} \frac{F_\Sigma}{2\pi} 
= - \chi(\Sigma_{g,n})  \ . 
\eeq

\subsection{The Form $E_4$ away from Punctures}
\label{sec_E4_in_the_bulk}

In this section we discuss the form $E_4$ in the bulk
of the Riemann surface, i.e.~away from punctures.
As per the general discussion of subsection \ref{sec_inclusion_of_punctures},
the 4-form $E_4$ is a closed form
invariant under the action of the structure group
of the $S^4$ fibration $M_6^{\rm bulk}$.
It is natural to exploit the decomposition \eqref{NW6_N2_decomposition}
and use a factorized $E_4$ of the form\footnote{By
writing down all possible terms compatible with $SO(2) \times SO(3)$
symmetry, one verifies that $E_4$ is given by $\cE_2 \wedge e_2^\Omega$
up to the exterior derivative of a globally-defined
3-form.
}
\beq \label{general_E4}
E_4 = \cE_2 \wedge e_2^\Omega \ . 
\eeq
Let us explain our notation. The form $e_2^\Omega$ 
is the global, $SO(3)$ invariant angular form
 for the $N_{SO(3)}$ bundle.
If we turn off the $N_{SO(3)}$ connection,
the form $e_2^\Omega$ reduces to a multiple
of the volume form on $S^2_\Omega$.
We normalize $e_2^\Omega$ according to
\beq
\int_{S^2_\Omega} e_2^\Omega = 1 \ .
\eeq
The explicit expression for $e_2^\Omega$
can be found in appendix \ref{appendix_global_ang_forms}.
The 2-form $\cE_2$ is closed and gauge-invariant.
We can write
\beq \label{bulk_cE2}
\cE_2 = N\, \bigg[ d\gamma \wedge \frac{D\phi}{2\pi}  - \gamma \, \frac{\cF}{2\pi}
\bigg] \ .
\eeq
The function $\gamma = \gamma(\mu)$ 
  is constrained by 
regularity conditions.
If we turn off all $N_{SO(2)}$ and $N_{SO(3)}$ connections,
$E_4$ becomes proportional to the volume form on an $S^4$.
Regularity of $E_4$ in the region
where $S^2_\Omega$ shrinks demands $\gamma(0) = 0$.
The normalization of $E_4$, \eqref{new_E4_normalization},
then fixes $\gamma(1) = 1$. To summarize,
\beq \label{gamma_values}
\gamma(0) =0 \ , \qquad \gamma(1) = 1 \ .
\eeq

Let us stress that, 
in our conventions, the integral of $E_4$
over any 4-cycle must be integrally quantized\footnote{We take the components of the 3-form potential $C_3$
to have mass dimension 3. The coupling of an M2-brane to $C_3$
is realized with a factor $e^{i \int C_3}$ in the path integral measure.
The quantity $\int_{\cC_4}  G_4/(2\pi)$ is an integer
for any 4-cycle $\cC_4$, up to the effects discussed in \cite{Witten:1996md},
which are not relevant in our setup.
The fact that the flux of $E_4$ is integrally quantized then
follows from \eqref{dG4_generic}.
}.
A trivial example of a flux quantization
condition is   \eqref{new_E4_normalization},
which simply states that $E_4$ counts the total number of M5-branes
in the stack.
A more interesting example of flux quantization is the relation
\beq \label{bulk_fluxes}
\int_{\Sigma_{g,n} \times S^2_\Omega} E_4 =  N \, \chi(\Sigma_{g,n})  \ ,
\eeq
which follows from \eqref{cF_on_Sigma}, \eqref{general_E4},
\eqref{bulk_cE2},   and
\eqref{gamma_values}.
In the  integral above, $\Sigma_{g,n} \times S^2_\Omega$
denotes 
the 4-cycle   obtained by combining the Riemann surface
and $S^2_\Omega$,  at fixed $\mu=1$, where $S^1_\phi$ shrinks.
Even though flux quantization conditions for $E_4$ are 
straightforward in the bulk of the Riemann surface,
they will play an essential role in section \ref{sec_punctures}
in constraining the local puncture geometries and flux configurations.

\subsection{The Bulk Contribution to Anomaly Inflow}

In the previous section we have
fixed a local expression for $E_4$ 
 in the bulk of the Riemann surface. 
We are therefore in a position to 
compute the bulk contribution $I_6^{\rm inflow}(\Sigma_{g,n})$ to
anomaly inflow, defined in \eqref{decomposing_inflow2}.
The derivation follows standard techniques,
and makes use of a result of Bott and Cattaneo \cite{bott1999integral}.
We refer the reader to appendix \ref{appendix_I6bulk}
for more details.
The result reads
\begin{align}
\label{I6inf_bulk_result}
I_6^{\rm inflow}(\Sigma_{g,n}) & = \frac {1}{12} \, \chi(\Sigma_{g,n}) \, N^3 \, \hat n^{\rm 4d} \, p_1(N_{SO(3)}) 
\nn \\
& - \frac{1}{48} \, \chi(\Sigma_{g,n}) \, N \, \hat n^{\rm 4d} \, \bigg[
p_1(TW_4) + p_1(N_{SO(3)}) - (\hat n^{\rm 4d})^2
\bigg] \ .
\end{align}
The notation $\hat n^{\rm 4d}$ was introduced in
\eqref{hatn_eq}. The quantities $p_1(TW_4)$,
$p_1(N_{SO(3)})$ are the first Pontryagin classes of 
the tangent bundle to external spacetime,
and the $N_{SO(3)}$ normal bundle,
respectively.

The quantities $\hat n^{\rm 4d}$ and $p_1(N_{SO(3)})$
are given in terms of the 4d Chern classes as
\beq \label{put_4d_Cherns}
\hat n^{\rm 4d} = 2 \, c_1^r \ , \qquad
p_1(N_{SO(3)}) = - 4 \, c_2^R \ ,
\eeq
where $c_1^r$ is a shorthand notation for the first
Chern class of the  4d   $U(1)_r$ R-symmetry bundle,
while $c_2^R$ is a shorthand notation for the second
Chern class of the  4d   $SU(2)_R$ R-symmetry bundle.
The bulk contribution to $I_6^{\rm inflow}$ then takes   the form
\beq \label{bulk_inflow_I6}
I_6^{\rm inflow}(\Sigma_{g,n})  =- \frac 16 \, \chi(\Sigma_{g,n}) \,  (4N^3 -N )  \, c_1^r \, c_2^R 
 - \frac{1}{24} \, \chi(\Sigma_{g,n}) \, N \, c_1^r \, \bigg[
p_1(TW_4)    - 4 \, (c_1^r)^2
\bigg] \ .
\eeq



\section{Introduction of Punctures}  
\label{sec_punctures}

In this section we 
discuss punctures and  analyze the properties of the local
geometries $X_6^\alpha$ introduced in section \ref{sec_inclusion_of_punctures}.
This analysis can be carried out separately for each puncture.
Therefore in what follows, we omit the puncture label $\alpha$,
and write $X_6$ for $X_6^\alpha$, $X_4$ for $X_4^\alpha$,
and so on.
We demonstrate that 
 the puncture data are encoded
in monopole sources for a suitable circle fibration,
and we analyze the form of $E_4$ in the vicinity of a puncture.

\subsection{Warm-up: Reformulation of a Non-puncture}

According to the strategy outlined in section 
\ref{sec_inclusion_of_punctures},
a non-trivial puncture can be described 
by removing a small disk $D$ from the Riemann
surface and replacing $D \times S^4$
with a new geometry $X_6$.
In order to gain insight into the properties of $X_6$
for   punctures,
we first analyze the case 
of a \emph{non-puncture}, i.e.~we set $X_6 = D\times S^4$
and   seek a reformulation of this 
trivial geometry that is best suited 
for generalizations to non-trivial spaces.
We show that 
$X_6 = D\times S^4$ can be recast
as an $S^2_\Omega$ fibration over a 4d space
$X_4$, which is in turn a circle fibration over $\mathbb R^3$.
We also provide a reformulation of the class $E_4$
that will prove beneficial in the discussion of
genuine punctures.

\subsubsection*{Geometry for the Non-puncture}
\label{sec_nonpuncture_geom}

\begin{figure}
\centering
\includegraphics[width = 11.5cm]{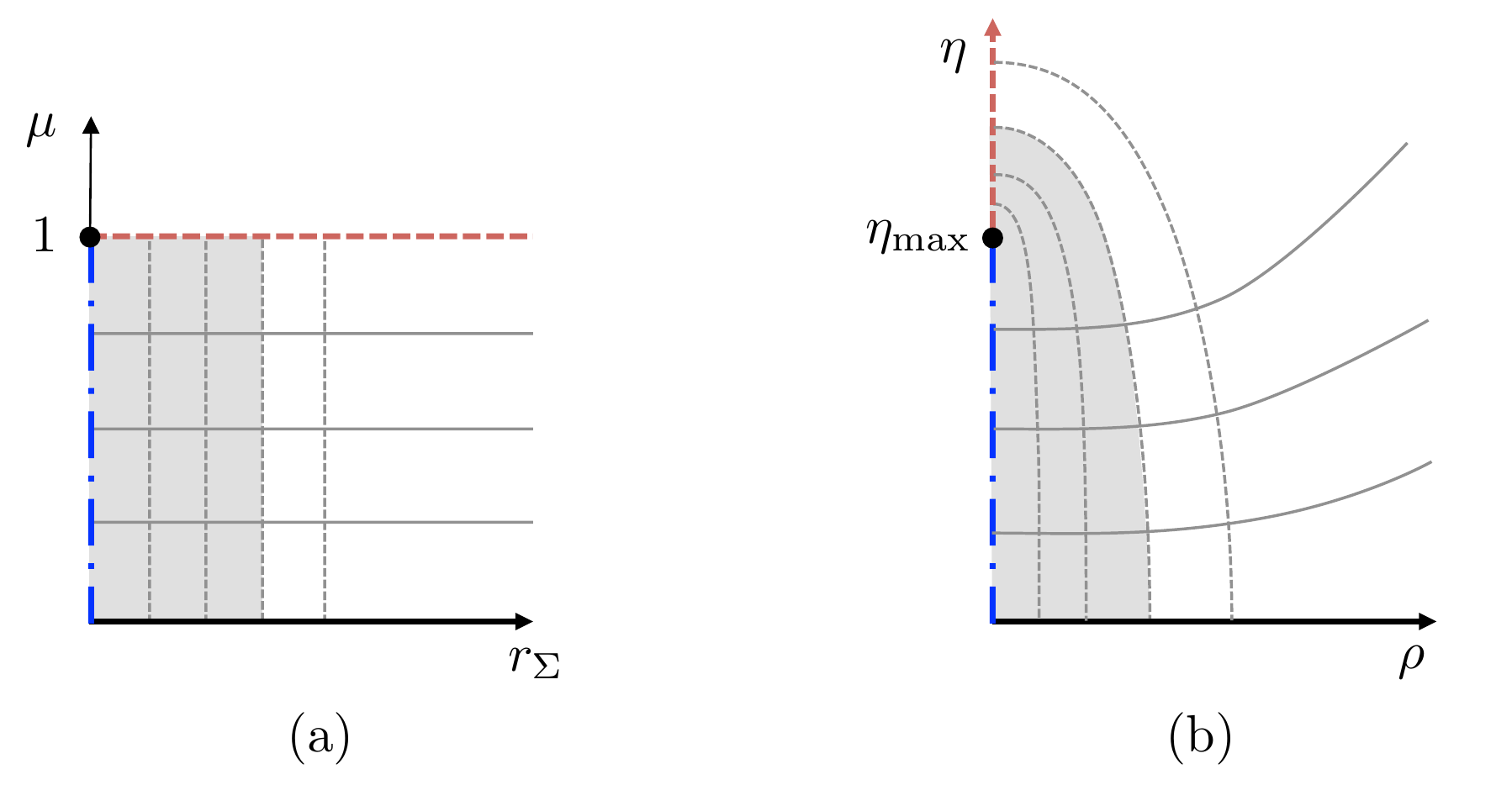}
\caption{The plot on the left depicts the $(r_\Sigma, \mu)$
strip, with $r_\Sigma$ on the horizontal axis,
and $\mu$ on the vertical axis. 
Lines of constant $\mu$ (solid, grey)
and lines of constant $r_\Sigma$ (dashed, grey)
are also included.
The plot on the right depicts the $(\rho, \eta)$
quadrant, 
with $\rho$ on the horizontal axis, and $\eta$
on the vertical axis. We include the image 
of lines of constant $\mu$ and $r_\Sigma$.
The shaded regions in both plots correspond to the subregion
 $r_\Sigma < r_0$, with $r_0$ constant.
}
\label{plots2}
\end{figure}

Our starting point is $X_6 = D \times S^4$. The disk $D$
is parametrized by standard polar coordinates $(r_\Sigma, \beta)$.
As usual, $S^4$ is realized as an $S^1_\phi \times S^2_\Omega$
fibration over the $\mu$ interval.
The line element on $X_6$ is simply
\beq \label{before_makeover}
ds^2 (X_6) = dr_\Sigma^2  + r_\Sigma^2 \, d\beta^2 + \frac{d\mu^2}{1-\mu^2}
+ (1-\mu^2)\,  D\phi^2 + \mu^2 \, ds^2(S^2_\Omega) \ , \qquad
D\phi = d\phi - A_\Sigma \ .
\eeq
We have recalled that $S^1_\phi$ is fibered over the Riemann surface
with a connection $A_\Sigma$.  For simplicity,
we have temporarily turned off all external connections.
The connection $A_\Sigma$
on the disk $D$ can be taken to be of the form
\beq \label{introducing_U}
A_\Sigma = U(r_\Sigma) \, d\beta \ , 
\eeq
where the function $U$
goes to zero as $r_\Sigma \rightarrow 0$
to ensure that $A_\Sigma$ is defined at the center of the disk.
The 2d space 
spanned by $r_\Sigma$
and $\mu$ is a half strip in the $(r_\Sigma, \mu)$
plane, described by
\beq
r_\Sigma \ge 0 \ , \qquad 0 \le \mu \le 1 \ ,
\eeq
see figure \ref{plots2} plot (a).
More precisely, the interior of the disk $D$ corresponds to a region
of the form $r_\Sigma < r_0$, with $r_0$ constant,
which is the shaded region
in figure \ref{plots2} plot (a).

Let us introduce a new angular coordinate $\chi$,
defined by
\beq \label{chi_def}
\chi = \phi + \beta \ .
\eeq
We can rewrite the line element \eqref{before_makeover} in the form
\begin{align} \label{after_makeover}
ds^2(X_6) & = ds^2(X_4) + \mu^2 \, ds^2(S^2_\Omega) \ , \nn \\
ds^2(X_4) & = 
ds^2(B_3)  
+ R^2_\beta \, D\beta^2  \ , \nn \\
ds^2(B_3) & = dr_\Sigma^2 + \frac{d\mu^2}{1-\mu^2}
+ \frac{r_\Sigma^2 \, (1 - \mu^2 )}{R^2_\beta} \, d\chi^2 \ ,
\end{align}
where we have introduced
\beq \label{non_puncture_details}
D\beta = d\beta - L \, d\chi \ , \qquad
L = \frac{1-\mu^2}{R^2_\beta} \ , \qquad
R^2_\beta = r^2_\Sigma + (1+U)^2 \, (1- \mu^2) \ .
\eeq
We have reinterpreted $X_6$ as an $S^2_\Omega$ fibration over 
a 4d space $X_4$. The latter is in turn written
as an $S^1_\beta$ fibration with connection $L \, d\chi$
over the 3d base space $B_3$ parametrized by $(r_\Sigma,\mu,\chi)$.
We can make the following   observations:
\begin{enumerate} [(i)]
\item The $S^2_\Omega$ shrinks on the locus
$(\mu = 0, r_\Sigma \ge 0)$,
 the thick black line 
in figure \ref{plots2} plot (a).
\item In the $(r_\Sigma,\mu)$ strip, the only
point where the 
$D\beta$ circle shrinks
is
$(r_\Sigma,\mu) = (0,1)$,
where the dot-dashed blue line and the  dashed red line meet in figure 
 \ref{plots2} plot~(a).
\item The $\chi$ circle \emph{in the 3d base space}
(which is specified in $X_4$ by $D\beta = 0$,
as opposed to $d\beta = 0$)
shrinks on the loci
$(r_\Sigma = 0 , 0 \le \mu \le 1)$
and $(\mu = 1, r_\Sigma \ge 0)$,
which correspond to the dot-dashed blue line
and the dashed red line in figure  \ref{plots2} plot (a), respectively.
\item The function $L$ is smooth in the interior of the $(r_\Sigma,\mu)$ strip.
Moreover, 
$L(r_\Sigma,\mu) = 1$
on the locus $(r_\Sigma = 0 , 0 \le \mu \le 1)$, i.e.~on the dot-dashed blue
line.
Similarly, 
$L(r_\Sigma,\mu) = 0$
on the locus $(\mu = 1, r_\Sigma \ge 0)$, i.e.~on the dashed red line.
\end{enumerate}

We see that $L$ has a discontinuity at the point
$(r_\Sigma, \mu) = (0,1)$ where the $D\beta$ circle shrinks.
The metric on $X_4$ near this point can be
modeled by a single-center Taub-NUT space,
showing that the $D\beta$ fibration has a  {monopole}
source. 
We write the Taub-NUT metric as
\begin{align} \label{TN_for_reference}
ds^2({\rm TN})  &= V^{-1} \, D\beta^2 + V \, (d\rho^2 + d\eta^2  + \rho^2 \, d\chi^2) \ ,
\nn \\
dD\beta &= - *_{\mathbb R^3} dV \ , \qquad
V = \frac{1/2}{\sqrt{\rho^2 + (\eta - \eta_{\rm max})^2}} \ ,
\end{align}
where
$\rho$, $\eta$, $\chi$ are standard cylindrical coordinates on $\mathbb R^3$.
The factor $1/2$ is related to the fact that, in our conventions,
$\beta$ has periodicity $2\pi$.
The coordinates $\rho$, $\eta$ are related to $r_\Sigma$, $\mu$ by
\beq \label{our_example}
\rho = r_\Sigma \, \sqrt{1-\mu^2} \ , \qquad
\eta = \eta_{\rm max} + \frac 12 (r_\Sigma^2  - 1 + \mu^2 ) \ ,
\eeq
as verified by comparing $ds^2(X_4)$ and $ds^2({\rm TN})$
near $(r_\Sigma, \mu) = (0,1)$, with $U = 0$ for simplicity.

The coordinate change \eqref{our_example} near 
$(r_\Sigma,\mu)=(0,1)$ is   a specific example of a general class
of maps with the qualitative features depicted in figure \ref{plots2} plot (b).
First of all, the $(r_\Sigma, \mu)$ half strip is
 mapped to the  quadrant in the $(\rho, \eta)$ plane
 with $\eta \ge 0$, $\rho \ge 0$.
Second of all, 
the thick black line is mapped to $\eta = 0$.
Finally,
the union of the dot-dashed blue line and the dashed
red line is mapped to the $\eta$ semi-axis.
The corner $(r_\Sigma, \mu)=(0,1)$ is mapped
to the point $\eta = \eta_{\rm max}$ on the $\eta$ axis.
The dot-dashed blue line is
mapped to the region $0 \le \eta \le \eta_{\rm max}$,
while the dashed red line is
mapped to $\eta \ge \eta_{\rm max}$.
Figure  \ref{plots2} plot (b) also shows
the   shaded region
corresponding to the interior of the disk $D$
in the new coordinates $(\rho,\eta)$\footnote{An
example
of a class of coordinate transformations from
$(r_\Sigma, \mu)$ to $(\rho, \eta)$ with the 
desired properties, different from \eqref{our_example}, is provided in \eqref{inverse_backlund}.}.

We have shown that the space $X_6 = D \times S^4$
can be reformulated as an $S^2_\Omega$ fibration
over a space $X_4$, which is in turn
a non-trivial $S^1_\beta$ fibration over
$\mathbb R^3$,
parametrized by cylindrical coordinates
$(\rho, \eta, \chi)$,
\beq \label{fibering_structure}
S^2_\Omega \hookrightarrow X_6 \rightarrow X_4 \ , \qquad
S^1_\beta \hookrightarrow X_4 \rightarrow \mathbb R^3 \ .
\eeq
In the above discussion, we have not included
the external connection $A_\phi$ for $\phi$.
If we turn $A_\phi$ on, \eqref{chi_def} indicates
that $d\chi$ should be replaced everywhere by
\beq \label{Dchi_def}
D\chi = d\chi - A_\phi \ .
\eeq
In particular, we must replace $d\chi$ with $D\chi$
inside $D\beta$, thus obtaining the quantity
\beq \label{tildeDbeta_def}
\widetilde {D \beta}= d\beta - L \, D\chi \ .
\eeq

\subsubsection*{The Form $E_4$ for the Non-puncture}

As explained in section \ref{sec_E4_in_the_bulk},
the form $E_4$ away from punctures takes the form
\eqref{general_E4} with $\cE_2$ given by \eqref{bulk_cE2}.
In light of the results of the previous section,
we seek a re-writing of $\cE_2$ in terms
of the 1-forms $D\chi$ and $\widetilde {D\beta}$
introduced in \eqref{Dchi_def}, \eqref{tildeDbeta_def}, respectively.
We are thus led to consider the ansatz
\beq \label{new_cE2}
\cE_2 = d\bigg[ Y \, \frac{D\chi}{2\pi} - W \, \frac{\widetilde { D \beta}}{2\pi} \bigg] \ ,
\eeq
where $Y$, $W$ are functions of $\rho$, $\eta$.
In order to match the above $\cE_2$ with
\eqref{bulk_cE2}  we have to set
\beq
Y (\rho, \eta) = N \, \gamma \, \Big[ 1 - (1+U) \, L \Big] \ , \qquad
W(\rho, \eta) = N \, \gamma \, (1+U) \ .
\eeq
Along the $\eta$ axis,  $Y$ is
piecewise constant,
\beq \label{Yvalues_non}
Y (0,\eta) = \left\{
\begin{array}{ll}
0 &\qquad  \text{for $0<\eta<\eta_{\rm max}$} \ , \\
N & \qquad \text{for $\eta>\eta_{\rm max}$} \ . 
\end{array}
\right.
\eeq
In particular, $Y$ is discontinuous at $\eta = \eta_{\rm max}$.
In contrast, $W$ is regular everywhere,
because both $\gamma$ and $U$ are regular
in the entire $(r_\Sigma,\mu)$ strip,
or equivalently the entire $(\rho,\eta)$ quadrant.
It is worth noting that
\beq \label{W_properties_non}
W(0,\eta_{\rm max}) = N \ .
\eeq
Finally, we observe that both $Y$ and $W$ 
vanish at $\eta = 0$ for any $\rho$,
\beq \label{vanish_at_the_bottom}
Y(\rho,0) = 0 \ , \qquad 
W(\rho,0)= 0 \ .
\eeq
This is  necessary to ensure regularity of $E_4$,
and follows from the fact that $Y$ and $W$ are 
 proportional to $\gamma$.
Recall the factorized form \eqref{general_E4}
and that $e_2^\Omega$ contains the volume
form on $S^2_\Omega$, which shrinks at $\eta = 0$.

Even though  $L$ and $Y$
are discontinuous 
along the $\eta$ axis at $\eta = \eta_{\rm max}$,
the form $\cE_2$
is smooth there.
To check this, 
we write $\cE_2$ in the form
\beq \label{cE2_explicit}
\cE_2 = (dY + W \, dL + L \, dW) \, \frac{D\chi}{2\pi} - (Y + L \, W) \frac{F_\phi}{2\pi}
- dW \, \frac{d\beta}{2\pi} \ .
\eeq
The terms $dY$ and $dL$ are a potential source of $\delta$ function
singularities,
\beq
dY \Big|_{\rho = 0} = (+N) \, \delta(\eta - \eta_{\rm max}) \, d\eta \ , \qquad
dL \Big|_{\rho = 0} = (-1) \, \delta(\eta - \eta_{\rm max}) \, d\eta \ ,
\eeq
where the prefactor of the $\delta$ function
is simply the jump of $Y$, $L$ across $\eta_{\rm max}$.
As we can see, the $\delta$ function singularities
cancel against each other in \eqref{cE2_explicit},
by virtue of  \eqref{W_properties_non}.
Notice also that the function $Y + L \, W$ is
continuous along the $\eta$ axis across the monopole
location.

\subsection{Local Geometry and Form $E_4$ for a Puncture}

We are now in a position to discuss 
the geometry and the form $E_4$   for non-trivial punctures.
In this section we show that all puncture data
are encoded in the fluxes of $E_4$ along the non-trivial
4-cycles of the geometry $X_6$.

\subsubsection*{Geometry for a Puncture}
\label{puncture_geometry}

The reformulation of the non-puncture geometry of
  section \ref{sec_nonpuncture_geom}
provides a natural starting point for the
construction of a genuine puncture geometry $X_6$,
and determines the correct gluing prescription of $X_6$ to
$M_6^{\rm bulk}$.
We utilize the same fibration structure \eqref{fibering_structure},
repeated here for the reader's convenience:
\beq  
S^2_\Omega \hookrightarrow X_6 \rightarrow X_4 \ , \qquad
S^1_\beta \hookrightarrow X_4 \rightarrow \mathbb R^3 \ .
\eeq
The space $\mathbb R^3$ is again parametrized
by cylindrical coordinates $(\rho, \eta, \chi)$,
and  
$S^2_\Omega$ shrinks at $\eta = 0$.
The $S^1_\beta$ fibration is of the form
\beq \label{Dbeta_defin}
D\beta = d\beta - L \, d\chi \ ,
\eeq
but with a more general $L(\rho,\eta)$ 
than in the non-puncture case.
In the base space $\mathbb R^3$,
the relevant portion of the $(\rho,\eta)$ quadrant
is a region analogous to the shaded region
in figure \ref{plots2} plot (b).
The unshaded region outside is identified
with the bulk of the Riemann surface.

In the non-puncture case, the $S^1_\beta$ fibration
has only one unit-charge monopole source
located at $\eta = \eta_{\rm max}$.
We now consider several monopoles
and allow for charges 
greater than one.
More precisely, we consider a configuration with $p \ge 1$
monopoles, located at $(\rho,\eta) = (0,\eta_a)$, 
$a = 1, \dots, p$. 
The last monopole location is identified with $\eta_{\rm max}$,
$\eta_p = \eta_{\rm max}$. For uniformity of notation,
we also define $\eta _0 := 0$.
The function $L(\rho,\eta)$ is piecewise constant along the $\eta$
axis,
with   jumps across each monopole location $\eta_a$.
We introduce the notation
\beq \label{ell_def}
L(0, \eta ) = \ell_a    \qquad
\text{for} \qquad  \eta_{a-1}<\eta<\eta_a   \ .
\eeq
We   also demand
\beq \label{L_at_infinity}
L( 0, \eta ) =0  \qquad
\text{for}  \qquad \eta>\eta_p = \eta_{\rm max}  \ .
\eeq
This condition
 guarantees that, along the $\eta$ axis for $\eta > \eta_{\rm max}$,
the $\chi$ circle \emph{in the base} (i.e.~setting $D\beta = 0$)
coincides with the $\phi$ circle. 
This allows us to glue the local puncture geometry
to the bulk of the Riemann surface in a straightforward way\footnote{It
is also interesting to explore more general possibilities,
in which the gluing involves a non-trivial identification
of circles. We briefly comment on 
this point in the conclusion.
}.

The charge of the monopole at $\eta = \eta_a$ is measured
by the discontinuity of the $L$ connection across $\eta = \eta_a$.
If $S^2_a$ denotes a small 2-sphere of radius $\epsilon$
surrounding $\eta = \eta_a$
in the base space spanned by $(\rho,\eta,\chi)$, we have
\beq   \label{P_and_ka}
- \int_{S^2_a} \frac{dD\beta}{2\pi} = 
\Big [ L \Big ]_{\eta = \eta_{  a} - \epsilon}
^{\eta = \eta_{  a} + \epsilon} =  \ell_{a+1} - \ell_{a} =: - k_a \ ,
\eeq
where the quantity $k_a$ is a non-negative integer\footnote{The
sign is inferred from the non-puncture case. 
Supersymmetry demands that all monopole charges carry the same sign.
}.
Combining \eqref{P_and_ka} and 
\eqref{L_at_infinity} we immediately derive
the important relation
\beq \label{P_values}
  \ell_a  = \sum_{b = a}^p k_{b}    \ .
\eeq
Since $k_a \ge 1$, the sequence $\{\ell_a\}_{a=1}^p$
is a decreasing sequence of positive integers.
As a final remark,  the non-puncture geometry
is recovered by setting $p = 1$, $k_1 = 1$.

\subsubsection*{Orbifold Singularities}
\label{orbifold_section}
A crucial aspect of the generalization from the non-puncture
to a genuine puncture is the possibility of a monopole charge
$k_a>1$. 
In analogy with the non-puncture case,
in the vicinity of $\eta= \eta_a$
the space $X_4$ is modeled by a
single-center Taub-NUT space ${\rm TN}_{k_a}$ with charge $k_a$.
The latter 
has an $\mathbb R^4/\mathbb Z_{k_a}$ orbifold singularity.
This singularity admits a minimal resolution
in terms of a collection of  $k_a-1$ copies of $\mathbb C \mathbb P^1$.
Let $\widetilde{\rm TN}_{k_a}$ denote the resolved Taub-NUT space.
In  $\widetilde{\rm TN}_{k_a}$,
each $\mathbb C \mathbb P^1$ has self-intersection number $-2$,
and the $\mathbb C \mathbb P^1$'s form a linear chain 
with intersection number $+1$ between distinct, neighboring 
$\mathbb C \mathbb P^1$'s.

In the resolved geometry $\widetilde{\rm TN}_{k_a}$, we use the symbol
$\widehat \omega_{a,I}$,
$I=1, \dots, k_a-1$ to denote the Poincar\'e dual
2-forms to the $\mathbb C \mathbb P^1$ cycles
resolving the singularity.
The forms $\widehat \omega_{a,I}$ satisfy
\beq \label{resolution_intersection}
\int_{ \widetilde{\rm TN}_{k_a} } \widehat  \omega_{a,I} \wedge
\widehat  \omega_{a,J} =
- \, C_{IJ}^{\mathfrak{su}(k_a)} \ ,
\eeq
where there is no sum over $a$ and the symbol $C_{IJ}^{\mathfrak{su}(k_a)}$ on the RHS denotes
the entries of the Cartan matrix of $\mathfrak{su}(k_a)$.

\subsubsection*{The Form $E_4$ for a Puncture}
\label{sec_puncture_E4}

Let us now discuss the structure of the form $E_4$ near a  
puncture. We assume the factorized form \eqref{general_E4}
and the ansatz \eqref{new_cE2} for $\cE_2$,
repeated here for convenience,
\beq \label{E4_repeated}
E_4 = \cE_2 \wedge e_2^\Omega + \dots \ , \qquad
\cE_2 = d\bigg[ Y \, \frac{D\chi}{2\pi} - W \, \frac{\widetilde { D \beta}}{2\pi} \bigg]
 \ ,
\eeq
where the dots represent terms associated to the flavor symmetry 
of the puncture, discussed in subsection \ref{including_flavor}.
In order to ensure regularity of $E_4$,
we must again demand that both $W(\rho,\eta)$ and $Y(\rho,\eta)$
vanish at $\eta = 0$, as in \eqref{vanish_at_the_bottom}.

In order to analyze the properties of $E_4$, we first have to
study the non-trivial 4-cycles in the puncture geometry $X_6$.
Below we construct two families of 4-cycles, denoted
$\{ \cC_a\}_{a=1}^{p}$ and $\{ \cB_a\}_{a=1}^p$.
As we shall see, regularity of $E_4$ at the monopole
locations implies that 
flux configurations are labelled by 
 a partition of $N$.

\begin{figure}
\centering
\includegraphics[width = 5.5cm]{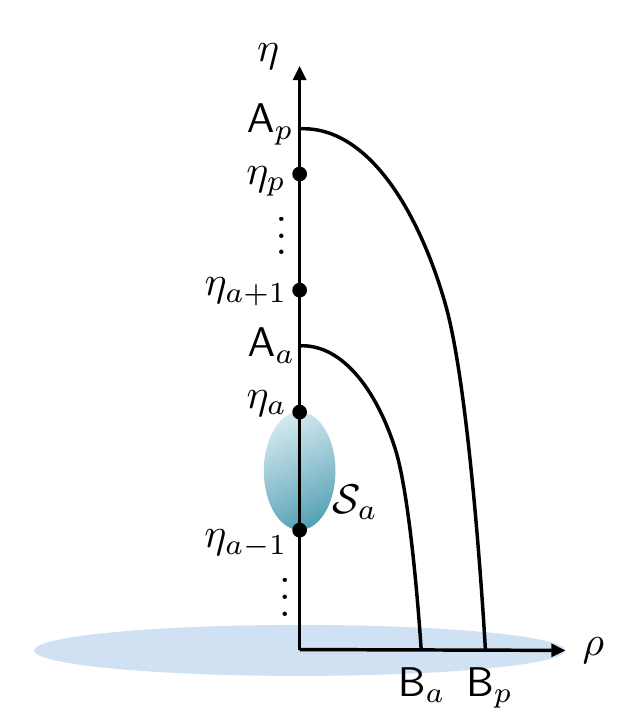}
\caption{
A generic profile of monopoles. The arcs $\mathsf A_a \mathsf B_a$
form part of the 4-cycle $\cC_a$. The bubble
denotes the 2-cycle $\cS_a$, which is part of the 4-cycle
$\cB_a$.
}
\label{all_cycles_fig}
\end{figure}

\paragraph{The 4-cycles $\{ \cC_a\}_{a=1}^p$.}
For each $a  = 1,\dots,p-1$,  the 4-cycle $\cC_a$
is constructed as follows.
In the $(\rho, \eta)$ quadrant,
pick an arbitrary point $\mathsf A_a$ in the interior of the interval $(\eta_a, \eta_{a+1})$
along the $\eta$ axis,
and an arbitrary point $\mathsf B_a$ with $\eta = 0$, $\rho >0$,
see figure \ref{all_cycles_fig}.
At the point $\mathsf A_a$, the $\chi$ circle \emph{in the base}, i.e.~at $D\beta = 0$,
is shrinking. At point $\mathsf B_a$, $S^2_\Omega$ is shrinking.
We thus obtain a 4-cycle with the topology
of an $S^4$ by combining the arc $\mathsf A_a \mathsf B_a$, the $\chi$ circle in the base,
and $S^2_\Omega$.
The same construction can be repeated by selecting
a point $\mathsf A_p$ along the $\eta$ axis in the region
$\eta > \eta_{\rm max}$. We denote the corresponding
4-cycle as $\cC_p$.
Crucially, by virtue of \eqref{L_at_infinity},
the $\chi$ circle in the base is nothing but $S^1_\phi$
for $\eta > \eta_{\rm max}$. It follows that
\beq \label{cCp_is_S4}
\cC_p \cong S^4 \ .
\eeq
This observation allows us to fix a uniform orientation
convention for all 4-cycles $\{ \cC_a\}_{a=1}^{p}$:
we must choose the convention
that ensures $\int_{\cC_p} E_4 = +N$,
see \eqref{new_E4_normalization}.

To compute the flux of $E_4$ through $\cC_a$,
with $a = 1, \dots, p-1$, 
we   enforce $D\beta = 0$ at point $\mathsf A_a$ by setting
$d\beta = \ell_{a+1} \, d\chi$. 
We   then 
 obtain
\beq \label{flux_Ca}
\int_{\cC_a} E_4 = \int_{\cC_a} e_2^\Omega \wedge
d\Big[ Y + (L-\ell_{a+1}) W \Big] \wedge \frac{d\chi}{2\pi} 
= - \Big[ Y + (L-\ell_{a+1}) W \Big]_{\mathsf A_a}^{\mathsf B_a}
= Y(\mathsf A_a)
\ .
\eeq
In the second step, we used $\int_{S^2_\Omega} e_2^\Omega=1$,
and we have recalled that $\chi$ has periodicity $2\pi$.
 In the final step,
we utilized $W(\mathsf B_a) = 0$, $Y(\mathsf B_a) = 0$ (which follow from \eqref{vanish_at_the_bottom})
and $L(\mathsf A_a) = \ell_{a+1}$
(which follows from \eqref{ell_def}).
While \eqref{flux_Ca} was derived under the assumption
$a  = 1, \dots, p-1$, it is   verified that it also holds
for $a = p$.

The computation \eqref{flux_Ca} deserves further comments.
First of all, since $\int_{\cC_a} E_4$ must be quantized
and the location of $\mathsf A_a$ inside the interval
$(\eta_{a}, \eta_{a+1})$ is arbitrary,
we learn that $Y(\rho,\eta)$ is piecewise constant along the $\eta$ axis.
We introduce the notation
\begin{align} \label{Y_values}
Y( 0, \eta ) &= y_a \in \mathbb Z  \qquad
\text{for} \qquad  \eta_{a}<\eta<\eta_{a+1}   \ , \qquad a = 1, \dots, p-1 \ , \nn \\
Y( 0, \eta ) &= y_p \in \mathbb Z  \qquad
\text{for} \qquad  \eta>\eta_{p} = \eta_{\rm max} \ . 
\end{align}
Notice that $y_0 = 0$, because $Y$ vanishes at $\eta = 0$.
Moreover, 
we can check that the orientation we chose in 
\eqref{flux_Ca} is consistent.
Indeed, \eqref{flux_Ca} holds for any choice of $p$ and $k_a$,
and in particular for the non-puncture.
In that case, \eqref{Yvalues_non} shows that $Y = N$
along the $\eta$ axis for $\eta > \eta_{\rm max}$.
We thus recover the expected relation $\int_{\cC_1} E_4 = +N$.

The identification  \eqref{cCp_is_S4}
provides the boundary condition 
\beq \label{yp_is_N}
y_p = N \ .
\eeq
For any puncture, supersymmetry requires  that
the flux of $E_4$
through all the $\cC_a$ 
carry the same sign. 
It follows that 
\beq
y_a > 0 \qquad \text{for} \qquad  a = 1, \dots, p-1 \ .
\eeq

\paragraph{The 4-cycles $\{ \cB_a\}_{a=1}^p$.}
For $a = 2, 3, \dots, p$, we can construct a 4-cycle $\cB_a$
as follows. Consider the interval $[\eta_{a-1}, \eta_a]$
on the $\eta$ axis. 
The circle $S^1_\beta$ 
shrinks   at the location of the monopole sources,
but has 
 finite size in the interior of 
$[\eta_{a-1}, \eta_a]$.
As a result, we can combine $S^1_\beta$ and 
$[\eta_{a-1}, \eta_a]$ and obtain a 2-cycle 
$\cS_a$ with the topology of an $S^2$,
depicted as a bubble in figure \ref{all_cycles_fig}.
The desired 4-cycle $\cB_a$ is then simply obtained
as $\cB_a = \cS_a\times S^2_\Omega$,
since $S^2_\Omega$ has finite size 
in the entirety of $[\eta_{a-1}, \eta_a]$.
We can also construct a 4-cycle $\cB_1$
by combining the interval $[0, \eta_1]$
with $S^1_\beta$ and $S^2_\Omega$.
In contrast with the case $a = 2, \dots, p$,
the $S^1_\beta$ circle is \emph{not}
shrinking at the endpoint $\eta= 0$.
However, $S^2_\Omega$ is shrinking there,
and therefore $\cB_1$ is  still a closed 4-cycle.

The flux of $E_4$ through the cycles $\{ \cB_a\}_{a=1}^p$
is computed from \eqref{E4_repeated}
by selecting the terms with one $D\beta$ factor,
\beq \label{Ba_fluxes}
\int_{\cB_a} E_4 =-  \int_{\cB_a} e_2^\Omega \wedge dW \wedge \frac{D\beta}{2\pi}
=W(0,\eta_a) - W(0, \eta_{a-1})
 \ .
\eeq
We have recalled that $e^\Omega_2$ integrates to $1$ over $S^2_\Omega$,
and that $\beta$ has periodicity $2\pi$. We have also chosen an orientation
for $\cB_a$.

To argue in favor of our orientation convention,
we specialize \eqref{Ba_fluxes} to the case of the non-puncture,
$p=1$, $\eta_1= \eta_{\rm max}$.
In that case, the cycle $\cB_1$ must be equivalent to 
$S^4$, since the latter is the only non-trivial 4-cycle in the non-puncture
geometry. From \eqref{W_properties_non}, \eqref{vanish_at_the_bottom},
we immediately see that the RHS of \eqref{Ba_fluxes}
evaluates to $+N$.

It follows from \eqref{Ba_fluxes}
that the  jumps in the values of $W$ between consecutive
monopole locations must be integers, by virtue
of $E_4$ flux quantization. Moreover,
supersymmetry demands that the flux of $E_4$
must have the same sign for all $\{ \cB_a \}_{a=1}^p$.
Consistency with the non-puncture case requires that this sign
must be positive.
In conclusion, we can write
\beq \label{W_values}
W(0, \eta_a) = w_a \ , \qquad
w_a - w_{a-1} \in \mathbb Z_+ \ ,\qquad
w_0 = 0\ ,
\eeq
where the last relation follows from  \eqref{vanish_at_the_bottom}.
Notice that   $\{w_a\}_{a=1}^p$
is an increasing sequence of positive integers.

\begin{figure}
\centering
\includegraphics[width = 11.cm]{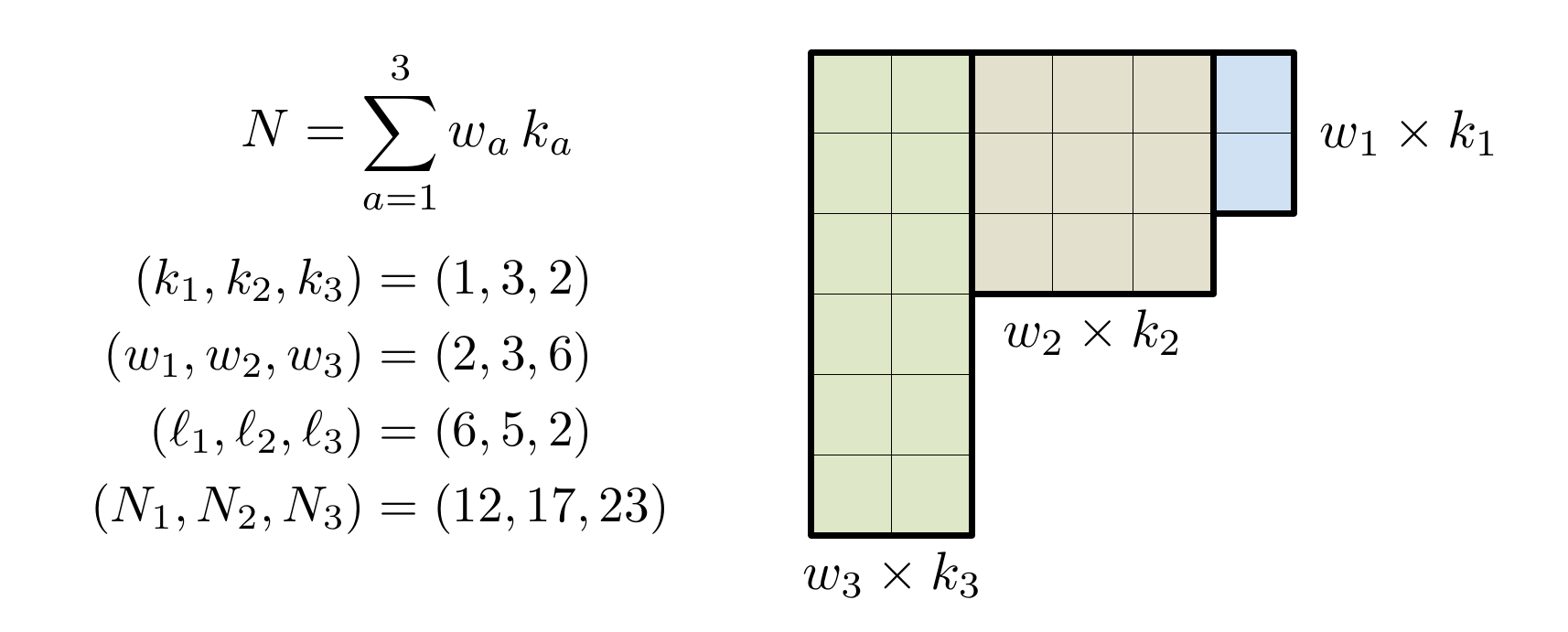}
\caption{
An example of a flux configuration for $N = 23$
and associated Young diagram.
The configuration has $p=3$ monopole sources
with prescribed $k_a$, $w_a$. We highlighted the 
decomposition of the Young diagram in  rectangular
blocks
of dimensions $w_a \times k_a$.
}
\label{young_from_flux}
\end{figure}

\paragraph{Regularity of $E_4$ and partition of $N$.}
The quantities $L$ and $Y$ are piecewise constant
along the $\eta$ axis, with   jumps 
at the location of the monopoles.
The total form $E_4$, however, must be regular
everywhere along the $\eta$ axis.
The terms $dL$ and $dY$ in \eqref{cE2_explicit}
 are a potential source of $\delta$ function
singularities in $\cE_2$, since
\beq
dY \Big|_{\rho = 0} = \sum_{a=1}^p (y_a-y_{a-1})\, \delta(\eta - \eta_a) \, d\eta \ , \qquad
dL \Big|_{\rho = 0} = \sum_{a=1}^p (\ell_{a+1} - \ell_a) \, \delta(\eta - \eta_{a}) \, d\eta \ .
\eeq
The normalization of each $\delta$ function at a given
$\eta_a$ is inferred from the jump of $Y$, $L$ across
$\eta_a$, see \eqref{Y_values}, \eqref{ell_def} respectively.
We can achieve a cancellation of each $\delta(\eta - \eta_a)$
term in \eqref{cE2_explicit} by demanding 
\beq  \label{Y_jumps}
0 = y_a - y_{a-1} + w_a \, (\ell_{a+1} - \ell_{a}) =
 y_a - y_{a-1} - w_a \, k_a   \ ,
\eeq
where in the last step we made use of \eqref{P_and_ka}.
We know from \eqref{vanish_at_the_bottom}
that $y_0 = 0$. As a result, we can use \eqref{Y_jumps}
to express the values of $y_a$ in terms of
$w_a$, $k_a$,
\beq \label{Y_values}
y_a =   \sum_{b=1}^{a} w_{b} \, k_{b}   \ .
\eeq
Moreover, we have also established that 
$y_p = N$, see \eqref{yp_is_N}.
Specializing \eqref{Y_values} to $a=p$
we thus obtain a crucial sum rule
for the flux data $w_a$, $k_a$,
\beq \label{flux_sum_rule}
N = \sum_{a=1}^p w_{a} \, k_{a} \ .
\eeq
Recall
that $\{w_a\}_{a=1}^p$
is an increasing sequence of positive integers, see \eqref{W_values}.
Moreover, all $k_a$ are integer and positive.
It follows that the relation \eqref{flux_sum_rule} 
defines a partition of $N$, which can be equivalently
encoded in a Young diagram.
Figure \ref{young_from_flux} exemplifies
the translation of \eqref{flux_sum_rule}
into a Young diagram, in the conventions
used throughout this work.

It is worth noting that, thanks to \eqref{Y_jumps}, the quantity $Y + W \, L$
is continuous along the $\eta$ axis\footnote{As explained
in appendix \ref{sec:GMappendix}, this quantity is the line charge density
in the Gaiotto-Maldacena puncture solutions \cite{Gaiotto:2009gz}.}.
 At the monopole location
$\eta = \eta_a$ it attains the value
\beq \label{Na_expression}
(Y+W \,L)(0,\eta_a) = 
N_a  = \sum_{b=1}^{a-1} w_{b} \, k_{b}
+ w_a \, \ell_a  =  \sum_{b=1}^a (w_{b} - w_{b-1}) \, \ell_{b}  \ .
\eeq
If we choose the last monopole $a = p$,
we can use $\ell_p = k_p$ (because 
$L$ is zero on the $\eta$ axis for $\eta > \eta_{\rm max}$)
and the sum rule \eqref{flux_sum_rule}
to infer $N_p = N$.


\subsubsection*{Flavor Symmetry}
\label{including_flavor}

In the case of the non-puncture, i.e.~$p=1$, $k_1=1$,
the space $X_4$ 
does not admit any non-trivial 2-cycles.
As soon as we consider more than one monopole source
and/or monopole charges greater than one, however,
the geometry $X_4$ contains non-trivial 2-cycles.
First of all, 
there are the 
2-cycles $\{ \cS_{a} \}_{a=2}^p$
introduced above \eqref{Ba_fluxes},
which have the topology of a 2-sphere and  are obtained by combining 
 the interval $[\eta_{a-1}, \eta_a]$
along the $\eta$ axis with the $D\beta$ fiber direction.
Let us stress once more that  
the $D\beta$ circle does \emph{not} shrink at $\eta_0 = 0$,
and therefore the first interval $[0,\eta_1]$ 
combined with $S^1_\beta$ does not yield
a 2-cycle.
The second class of 2-cycles in $X_4$
 is the collection of resolution $\mathbb C \mathbb P^1$'s
at each monopole source with $k_a > 1$, introduced
at the end of section \ref{puncture_geometry} above.

The existence of non-trivial 2-cycles in $X_4$
allows us to include additional terms in $E_4$.
The total $E_4$ thus reads
\beq \label{flavor_E4}
E_4 = \cE_2 \wedge e_2^\Omega
+ \sum_{a =2}^p \frac{F_{a}}{2\pi} \wedge \omega_{a}
+ \sum_{a = 1}^p \sum_{I=1}^{k_a-1}\frac{\widehat F_{a,I}}{2\pi} \wedge \widehat \omega_{a,I} \ ,
\eeq
where $\cE_2$ is as in \eqref{E4_repeated}.
The quantities $F_a$, $\widehat F_{a,I}$
are field strengths of 4d external connections.
The 2-form $\omega_a$ is the Poincar\'e dual in $X_4$
of the 2-cycle $\cS_a$, while the 2-forms $\widehat \omega_{a,I}$ are the Poincar\'e
duals of the resolution $\mathbb C \mathbb P^1$'s at each monopole
with $k_a \ge 2$. (The sum over $I$ is understood to be zero
if $k_a= 1$.)
The 4d connections $F_a$, $\widehat F_{a,I}$ in \eqref{flavor_E4}
are interpreted as background gauge fields
for the flavor symmetry associated to the puncture.
More precisely, \eqref{flavor_E4} captures the
Cartan subalgebra of the full flavor symmetry group
\beq \label{flavor_group}
G_F = S \bigg[ \prod_{a=1}^p U(k_a) \bigg] \ .
\eeq
The connections $\widehat F_{a,I}$ 
are in one-to-one correspondence
with the Cartan generators of the $SU(k_a)$ factor in $G_F$,
while the $F_a$ correspond to the remaining $U(1)$ factors.

The states associated to the non-Cartan generators of $G_F$
are not visible in the supergravity approximation,
since they originate from M2-branes wrapping the resolution
$\mathbb C \mathbb P^1$'s. For the purpose of computing 't Hooft anomaly coefficients,
however, the form $E_4$ contains all necessary information.



\section{Puncture Contributions to Anomaly Inflow}  
\label{sec_puncture_anomaly}

As explained in section \ref{sec_inclusion_of_punctures},
the contribution of the $\alpha^{\rm th}$ puncture
to the total inflow anomaly polynomial 
$I_6^{\rm inflow}(P_\alpha)$
is given by \eqref{decomposing_inflow2},
with $\cI_{12}$ given by \eqref{I12_def}.
In this section we compute the integral in 
\eqref{decomposing_inflow2}, considering the two terms in $\cI_{12}$ in turn.
For notational convenience, we  suppress the puncture label $\alpha$
throughout the rest of this section.

\subsection{Computation of the $(E_4)^3$ Term}
\label{E4cube_subsec}
The total expression for the form $E_4$ near a puncture
is given in \eqref{flavor_E4}, with   $\cE_2$ as in \eqref{E4_repeated}.
Our task is to identify the terms in $(E_4)^3$ that saturate the integral over 
the 6d space $X_6$, which is an $S^2_\Omega$ fibration over $X_4$,
see \eqref{fibering_structure}.
The Bott-Cattaneo formula,  
reviewed in appendix \ref{appendix_bott_cattaneo}, implies
\beq
\int_{S^2_\Omega} (e_2^\Omega)^3 =  \frac 14 p_1(N_{SO(3)})  \ , 
\qquad
\int_{S^2_\Omega} e_2^\Omega = 1 \ ,
\eeq
while even powers of $e_2^\Omega$ integrate to zero.
It follows that
\beq \label{E4cube_intermediate}
\int_{X_6} (E_4)^3 = \frac 14 p_1(N_{SO(3)})  \, \int_{X_4} (\cE_2)^3
+ 3\,  \int_{X_4}    \cE_2 \wedge \bigg[
\sum_{a =2}^p \frac{F_{a}}{2\pi} \wedge \omega_{a}
+ \sum_{a = 1}^p \sum_{I=1}^{k_a-1}\frac{\widehat F_{a,I}}{2\pi} \wedge \widehat \omega_{a,I}\bigg]^2 \ .
\eeq
To proceed, we isolate the terms in $(\cE_2)^3$ that saturate the integration
over $X_4$,  
\beq
\int_{X_4} (\cE_2)^3 = -3  \int_{X_4}   d \Big[ (Y + W \, L)^2 \Big] \wedge 
dW  \wedge \frac{D\chi}{2\pi} \wedge \frac{\widetilde {D\beta }}{2\pi} \wedge   \frac{F_\phi}{2\pi}  \ .
\eeq
The integration over the angles $\chi$, $\beta$
is readily performed. (Recall that they both have periodicity $2\pi$.)
The integral of the 2-form 
$d [ (Y + W \, L)^2 ] \wedge 
dW$ on the $(\rho, \eta)$ plane
is discussed in detail in appendix \ref{integral_for_E4cube}.
Combining all elements,
we arrive at
\beq
\int_{X_4} (\cE_2)^3 = 
-   \frac{F_\phi}{2\pi} \,  \sum_{a=1}^p \bigg[
  2 \, \ell_a^2 \, (w_a^3 - w_{a-1}^3)
+ 3 \, \ell_a \, (N_a - w_a \, \ell_a) \, (w_a^2 - w_{a-1}^2) 
\bigg]  \ .
\eeq

Let us now turn to the second $X_4$ integral in 
\eqref{E4cube_intermediate}. In this case,
integration over $X_4$ is saturated by considering terms
quadratic in the 2-forms $\omega_a$, $\widehat \omega_{a,I}$,

\begin{align} \label{some_step}
\int_{X_6} (E_4)^3 & = 
-  3\,   \frac{ F_\phi }{ 2\pi} \wedge \bigg[ \sum_{a,b=1}^p \sum_{I=1}^{k_a-1} \sum_{J=1}^{k_b-1}
\cK_{(a,I),(b,J)} \, 
 \frac{ \widehat F_{a,I} }{ 2\pi} \wedge
  \frac{ \widehat F_{b,J} }{ 2\pi}  
  \nn \\
&+ \sum_{a,b=2}^p  \cK_{a,b}  \,
 \frac{   F_{a} }{ 2\pi} \wedge
  \frac{   F_{b} }{ 2\pi}  
       +2 \, \sum_{a=2}^p \sum_{b=1}^p \sum_{J=1}^{k_b-1} 
\cK_{a,(b,J)} \,
 \frac{   F_{a} }{ 2\pi} \wedge
  \frac{ \widehat F_{b,J} }{ 2\pi}  \bigg]+ \dots \ ,
\end{align}
where the coefficients are 
\begin{align} \label{Z_coeffs}
\cK_{(a,I), (b,J)} &= \int_{X_4}(Y + W \, L)  \,  
  \widehat \omega_{a,I} \wedge  \widehat \omega_{b,J}  \ , \nn \\
 \cK_{a,b} &= \int_{X_4}(Y + W \, L)  \,  
    \omega_{a} \wedge    \omega_{b}  \ ,  \nn \\
  \cK_{a, (b,J)} &= \int_{X_4}(Y + W \, L)  \,  
    \omega_{a} \wedge  \widehat \omega_{b,J}  \ . 
\end{align}
We have  used the fact that the only relevant
part of $\cE_2$ is the one with legs along external spacetime,
$ -(Y + W \, L) F_\phi /(2\pi)$.

The coefficients $\cK_{(a,I),(b,J)}$ 
are computed as follows.
The 2-forms $\widehat \omega_{a,I}$ are 
associated to the resolution $\mathbb C \mathbb P^1$'s 
of the orbifold singularity at the $a^{\rm th}$ monopole.
It follows that $\cK_{(a,I),(b,J)}$
is only non-zero for $a=b$. As a result, the quantity
$Y+WL$ is evaluated at $(\rho,\eta) = (0,\eta_a)$,
and gives a factor $N_a$ by virtue of \eqref{Na_expression}.
The integration over $X_4$ reduces
to an integration over the resolved orbifold
$\widetilde {\rm TN}_{k_a}$
and is performed using \eqref{resolution_intersection}.
We thus have
\beq
\cK_{(a,I), (b,J)} = - \delta_{a,b} \, N_a \, C_{IJ}^{\mathfrak {su}(k_a)}    \ .
\eeq
A computation of the coefficients $\cK_{a,b}$ and $\cK_{a,(b,J)}$
in \eqref{Z_coeffs}
requires full control over the intersection pairing 
among the 2-cycles $\cS_a$ and the resolution
$\mathbb C \mathbb P^1$'s, as well as 
over the
normalization of the 2-forms $\omega_a$.
We refrain from a discussion of these coefficients.

Let us summarize the final result of the computation
of this subsection, 
using \eqref{put_4d_Cherns}
to express $\hat n^{\rm 4d} = F_\phi/(2\pi)$ and $p_1(N_{SO(3)})$ in terms
of 4d Chern classes,
\begin{align}
\int_{X_6} (E_4)^3 &= 
2 \,  c_1^r \, c_2^R
 \, 
\sum_{a=1}^p \bigg[
  2 \, \ell_a^2 \, (w_a^3 - w_{a-1}^3)
+ 3 \, \ell_a \, (N_a - w_a \, \ell_a) \, (w_a^2 - w_{a-1}^2) 
\bigg]  
  \nn \\
& + 6 \, c_1^r \, \sum_{a=1}^p  N_a \sum_{I,J  = 1}^{k_a-1} C_{IJ}^{\mathfrak{su}(k_a) } \, 
\frac{\widehat F_{a,I}}{2\pi} \wedge \frac{\widehat F_{a,J}}{2\pi} 
- 6 \, c_1^r \, \sum_{a,b=2}^p \cK_{a,b} \,
\frac{ F_{a}}{2\pi} \wedge \frac{  F_{b}}{2\pi}  \nn \\
& - 12 \, c_1^r \, \sum_{a=2}^p \sum_{b=1}^p \sum_{J=1}^{k_b-1}
\cK_{a,(b,J)}  \,
\frac{  F_{a}}{2\pi} \wedge \frac{\widehat F_{b,J}}{2\pi}  \ .
\end{align} 

\subsection{Computation of the $E_4 \wedge X_8$ Term}
\label{E4X8_subsec}

Recall from section \ref{orbifold_section} that
the puncture geometry $X_4$ has an $\mathbb R^4/\mathbb Z_{k_a}$
orbifold singularity at the location of each monopole of charge $k_a \ge 2$.
The singularity is modeled by a single-center Taub-NUT space
${\rm TN}_{k_a}$, which can be resolved to $\widetilde {\rm TN}_{k_a}$.
We use the notation $\widetilde X_4$ for the space
obtained from $X_4$ by resolving \emph{all} its orbifold
singularities.

With this notation, the relevant decomposition of the 11d tangent bundle,
restricted to the brane worldvolume, 
is
\beq \label{decomposition_puncture}
TM_{11} |_{W_6} = TW_4 \oplus N_{SO(3)} \oplus T\widetilde X_4 \ . 
\eeq
The above expression is motivated by the fact that the resolved
space $\widetilde X_4$ is a local model of the cotangent bundle
to the Riemann surface in the vicinity of the puncture.

Let $\lambda_1$, $\lambda_2$ denote the Chern roots of $T\widetilde X_4$.
Since $c_1(T\widetilde X_4) = 0$, we can write
\beq
\lambda_{1} = - \lambda_{2} = : \lambda \ .
\eeq
In our  geometry, the $U(1)$ associated to the $\chi$ circle
is gauged with the 4d
connection $A_\phi$. In order to account for this fact,
we shift the Chern roots of $T\widetilde X_4$,
\beq \label{lambda_shift}
\lambda_{1}^{\rm tot} := \lambda + \tfrac 12 \, \hat n^{\rm 4d}  \ , \qquad
\lambda_{2}^{\rm tot} := -\lambda + \tfrac 12 \, \hat n^{\rm 4d} \ ,
\eeq
where $\hat n^{\rm 4d}$ is the spacetime component of the
Chern root of $N_{SO(2)}$ introduced in \eqref{hatn_eq}.
We see from \eqref{lambda_shift} that it is the sum of Chern roots
$\lambda_1 + \lambda_2$
that is shifted with $+ \hat n^{\rm 4d}$.
This is due to the definition of the angle $\chi$ in terms
of $\beta$, $\phi$---see \eqref{chi_def}.
We can now   compute the shifted
Pontryagin classes for $T \widetilde X_4$, including the contribution
from the gauging with $\hat n^{\rm 4d}$,
\begin{align}
   p_1(T \widetilde X_4) ^{\rm tot} & =   (\lambda_{1}^{\rm tot})^2
+ (\lambda_{2}^{\rm tot})^2= p_1(T\widetilde X_4)
+   \tfrac 12 \,  (\hat n^{\rm 4d})^2 \ , \nn \\
   p_2(T \widetilde X_4) ^{\rm tot} & =  (\lambda_{1}^{\rm tot})^2
  \, (\lambda_{2}^{\rm tot})^2 = 
 - \tfrac 14 \,  p_1(T \widetilde X_4) \, (\hat n^{\rm 4d})^2 \ ,
\end{align}
where $p_1(T\widetilde X_4)$ is the first Pontryagin class
of $T\widetilde X_4$ before the 4d gauging is turned on.
Using \eqref{decomposition_puncture}, \eqref{I8def}, and standard formulae for Pontryagin classes
\eqref{eq:psum},
we compute
\begin{align}
X_8 &  = - \frac{1}{96} \, \Big[ p_1(TW_4) + p_1(N_{SO(3)}) 
-    (\hat n^{\rm 4d})^2  \Big]  {p_1(T\widetilde X_4)}  + \dots \ .
\end{align}
We have selected the terms with one $p_1(T \widetilde X_4)$,
with the dots representing the remaining terms,
which will not be important for the following discussion.

We are now in a position to integrate $E_4 \wedge X_8$
over $X_6$. 
The integral in the directions of $\widetilde X_4$
is saturated by $p_1(T\widetilde X_4)$,
while the integral on $S^2_\Omega$ is saturated
by $e_2^\Omega$ in the term $\cE_2 \wedge e_2^\Omega$ in $E_4$.
It follows that 
\beq
\int_{X_6} E_4 \wedge X_8 = \frac{1}{48} \,
\, c_1^r
\,  \Big[ p_1(TW_4) + p_1(N_{SO(3)}) 
-   (\hat n^{\rm 4d})^2  \Big] \, \int_{\widetilde X_4} (Y +  W \, L) \,  {p_1(T\widetilde X_4)}  \ .
\eeq
We have already performed the integral over $S^2_\Omega$,
and we have selected the only piece of $\cE_2$ which is relevant,
i.e.~the part with   $F_\phi =2\pi \hat n^{\rm 4d}$.
The integral over $\widetilde X_4$ localizes onto the
positions $\eta= \eta_a$ of the monopoles,
\beq
 \int_{\widetilde X_4} (Y +  W \, L) \,  {p_1(T\widetilde X_4)}
= \sum_{a=1}^p N_a \, \int_{\widetilde {\rm{TN} }  _{k_a}} p_1( T \widetilde {\rm{TN} }  _{k_a} ) \ .
\eeq
We exploited the fact that the quantity $Y + W \,L$
takes the value $N_a$ at $(\rho,\eta) = (0,\eta_a)$, see \eqref{Na_expression}.
The integrals of the  individual classes  $p_1( T \widetilde {\rm{TN} }  _{k_a} )$
are  evaluated making use of the results of \cite{Gibbons:1979gd}
for ALF resolutions of $\mathbb R^4/\mathbb Z_{k_a}$\footnote{Equation
(12) in 
\cite{Gibbons:1979gd} gives the Euler characteristic $\chi$ 
for a generic ALF space based on $\mathbb R^4/\Gamma$.
Exploiting self-duality of curvature,
specializing to $\Gamma = \mathbb Z_s$, 
using equation (23) in \cite{Gibbons:1979gd},
and comparing with the value of $\chi$ given in equation (33) in \cite{Gibbons:1979gd},
one reads out the integral of $p_1( T \widetilde {\rm{TN} }  _{k_a} )$. 
},
\beq
\int_{\widetilde {\rm{TN} }  _{k_a}} p_1( T \widetilde {\rm{TN} }  _{k_a} ) = 2 \, k_a \ .
\eeq 
In conclusion, we obtain
\beq
\int_{X_6} E_4 \wedge X_8 = \frac{1}{24} \sum_{a=1}^p  N_a \, k_a \,   c_1^r
\,  \Big[ p_1(TW_4) - 4 \, c_2^R  
- 4 \, (c_1^r)^2  \Big]    \ ,
\eeq
where we have expressed the result in terms of $c_1^r$, $c_2^R$ using 
 \eqref{put_4d_Cherns}.



\section{Comparison with CFT Expectations}  
\label{sec_comparison_with_CFT}

In this section we first summarize the total result for the
inflow anomaly polynomial, and we then prove that it
matches with the CFT expectation.

\subsection{Summary of Inflow Anomaly Polynomial}

We can assemble the contribution
$I_6^{\rm inflow}(P_\alpha)$ of the $\alpha^{\rm th}$ puncture 
to the inflow anomaly polynomial,
making use of \eqref{decomposing_inflow2} and the findings of the previous sections.
The result reads
\begin{align}
I_6^{\rm inflow}(P_\alpha) &= (n_v - n_h)^{\rm inflow}(P_\alpha) \, \bigg[
\frac 13 \, (c_1^r)^3 - \frac{1}{12} \, c_1^r \, p_1(TW_4)
\bigg]
- n_v^{\rm inflow}(P_\alpha) \, c_1^r \, c_2^R  \nn \\
& + I_6^{\rm inflow, flavor}(P_\alpha) \ ,
\end{align}
where the anomaly coefficients are given in terms of the quantized flux data as
\begin{align}
(n_v - n_h)^{\rm inflow}(P_\alpha) &= \frac 12\,  \sum_{a=1}^p N_a \, k_a \ , \nn
  \\
n_v^{\rm inflow}(P_\alpha) &= 
 \sum_{a=1}^p \bigg[
 \frac 2 3  \, \ell_a^2 \, (w_a^3 - w_{a-1}^3)
+   \ell_a \, (N_a - w_a \, \ell_a) \, (w_a^2 - w_{a-1}^2) 
- \frac 16 \, N_a \, k_a
\bigg] 
 \ , \label{eq:inflowans} \nn \\ 
I_6^{\rm inflow, flavor} (P_\alpha)& = 
  c_1^r \bigg[
  - \sum_{a=1}^p  N_a \sum_{I,J  = 1}^{k_a-1} C_{IJ}^{ \mathfrak{su}(k_a) } \, 
\frac{\widehat F_{a,I}}{2\pi} \wedge \frac{\widehat F_{a,J}}{2\pi} 
+ \, \sum_{a,b=2}^p \cK_{a,b} \,
\frac{ F_{a}}{2\pi} \wedge \frac{  F_{b}}{2\pi}  \nn \\
&  \qquad  \;\;\, +2 \, \sum_{a=2}^p \sum_{b=1}^p \sum_{J=1}^{k_b-1}
\cK_{a,(b,J)}  \,
\frac{  F_{a}}{2\pi} \wedge \frac{\widehat F_{b,J}}{2\pi} 
\bigg] \ .  
\end{align}
The coefficients $\cK_{a,b}$, $\cK_{a,(b,J)}$ in
 $I_6^{\rm inflow, flavor} (P_\alpha)$ 
are defined in \eqref{Z_coeffs}.
 A minor comment about our notation is in order.
We have reinstated the puncture label $\alpha$
on the LHS's of the above equations.
Strictly speaking, each puncture comes with its
local data, and on the RHS's we should write
$p^\alpha$, $k_a^\alpha$, $\ell_a^\alpha$, and so on.
We prefer to omit the label $\alpha$ from the RHS's
of the above relations in order to avoid cluttering the expressions.

In the piece related to flavor symmetry,
we expect an enhancement of the first 
term to the second Chern class of the 
full non-Abelian $SU(k_a)$ factor
in the flavor symmetry group,
\beq
  - c_1^r \, \sum_{a=1}^p  N_a \sum_{I,J  = 1}^{k_a-1} C_{IJ}^{ \mathfrak{su}(k_a) } \, 
\frac{\widehat F_{a,I}}{2\pi} \wedge \frac{\widehat F_{a,J}}{2\pi} 
\rightarrow - 2  \, \sum_{a=1}^p 2 \, N_a \, c_1^r \, c_2(SU(k_a))  \ .
\eeq
The corresponding flavor central charge is
\beq
k_{  {SU}(k_a)}^{\rm inflow} = - 2 \, N_a \ .\label{eq:flavorss}
\eeq

For the sake of completeness, we also restate the bulk contribution of the Riemann surface to the anomalies: 
\begin{align}
(n_v - n_h)^{\rm inflow}(\Sigma_{g,n}) &=  \frac{1}{2}\chi(\Sigma_{g,n}) N\ , \label{eq:bulk1}  \\
n_v^{\rm inflow}(\Sigma_{g,n}) &=   \frac{1}{6} \chi(\Sigma_{g,n})  (4N^3-N)\ . \label{eq:bulk2}
\end{align}
We would now like to compare these expressions with the anomalies of the 4d SCFT. Our results are  summarized in \eqref{eq:resultbulk1}-\eqref{eq:result3}.

 \subsection{Anomalies of the $\CN=2$ Class $\CS$ SCFTs}
 \label{sec:anomaliesn2}

The anomaly polynomial of any 4d
$\cN = 2$ SCFT with flavor symmetry $G_F$
can be written in the form
\beq \label{eq:CFT_general_anomaly}
I_6^{\rm CFT} = (n_v - n_h) \, \bigg[
\frac 13 \, (c_1^r)^3 - \frac{1}{12} \, c_1^r \, p_1(TW_4)
\bigg]
- n_v \, c_1^r \, c_2^R 
- k_G \, c_1^r \, {\rm ch}_2(G_F) \ .
\eeq
This structure follows from the $\CN=2$ superconformal algebra \cite{Kuzenko:1999pi}.
Here, ${\rm ch}_2(G_F)$ is the 2-form part of the Chern character for $G_F$;
for instance, ${\rm ch}_2(SU(m)) = -c_2(SU(m))$ (see appendix \ref{appendix_classes}).
The flavor central charge is defined in terms of the $G_F$ generators $T^a$ as 
	\ba{
	k_{G_F} \delta^{ab} = -2 \tr R_{\CN=2}T^aT^b.
	}
The parameters $n_v$ and $n_h$ correspond to the number of vector multiplets and hypermultiplets respectively when the theory is free, and otherwise can be regarded as an effective number of vector multiplets and hypermultiplets. These are related to the SCFT central charges as $a=\frac{1}{24}(5n_v+n_h)$, and $c=\frac{1}{12} (2n_v+n_h)$.

An $\CN=2$ theory of class $\CS$ has two contributions to their anomalies, which we denote in terms of the $n_v$ and $n_h$ parameters as 
	\ba{
	n_v^{\text{CFT}} = n_v^{\text{CFT}}(\Sigma_{g,n})+ \sum_{\alpha=1}^n n_v^{\text{CFT}}(P_\alpha)\ ,\qquad n_h^{\text{CFT}} = n_h^{\text{CFT}}(\Sigma_{g,n}) + \sum_{\alpha=1}^n n_h^{\text{CFT}}(P_\alpha)\ . \label{eq:totsn}
	}
The bulk terms are proportional to the Euler characteristic $\chi$ of the Riemann surface,  
	\ba{
	(n_v-n_h)^{\text{CFT}}(\Sigma_{g,n}) &= - \frac{1}{2} \, \chi(\Sigma_{g,n}) \,  (N-1),\   \label{eq:nvnhbulk}\\
	n_v^{\text{CFT}}(\Sigma_{g,n} ) &=  - \frac{1}{2} \, \chi(\Sigma_{g,n}) \, \left( \frac{4}{3} N^3- \frac{1}{3} N - 1\right)  \ .\label{eq:nvbulk}
	}
These were computed in \cite{Benini:2009mz,Alday:2009qq} by integrating the 6d (2,0) anomaly polynomial over the Riemann surface without punctures.  
The remaining terms  in \eqref{eq:totsn} depend on the local puncture data, which we will now review.

 A regular $\CN=2$ puncture is labeled by an embedding $\rho:\mathfrak{su}(2)\to \mathfrak{g}$.  For $\mathfrak{g}=A_{N-1}$, $\rho$ is 
 one-to-one with a partition of $N$, encoded in  a Young diagram with $N$ boxes. Consider a Young diagram with $\widetilde{p}$ rows  of length $\widetilde{\ell}_{i}$, with $i=1,\dots,\widetilde{p}$.  The partition is given as
 	\ba{
	N = \sum_{i=1}^{\widetilde{p}} \widetilde{\ell}_{i}\ .
	} 
A puncture corresponding to this partition  contributes a flavor symmetry $G_F$ to the 4d CFT, where $G_F$ is the commutant of the embedding $\rho$, 
	\ba{
	G_F = S\left[ \prod_{i=1}^{\widetilde p} U(\widetilde{k}_i) \right]. \label{eq:flavorsymm1}
	}
 The quantities $\widetilde k_i$ are defined as  
\ba{
	\widetilde{k}_{i} &= \widetilde{\ell}_{i} - \widetilde{\ell}_{i+1} \ ,\qquad \widetilde{\ell}_{\widetilde{p}+1}\equiv 0\ ,
	\qquad
	\widetilde{\ell}_{i} = \sum_{j=i}^{\widetilde{p}} \widetilde{k}_{j}  \ . \label{eq:k}
}
In order to write down $n_{v,h}^{\text{CFT}}(P_\alpha)$
it is also useful to introduce the notation 
\ba{
	\widetilde{\ell}_{i} = \widetilde{N}_{i} -  \widetilde{N}_{i-1} \ , 
	\qquad
	 \widetilde{N}_{\widetilde{p}} = \widetilde{N}_{\widetilde{p}+1}=N \ ,  
	 \qquad
	\widetilde{N}_{i} &=  \sum_{j=1}^{i} \widetilde{\ell}_{j}
	= \sum_{j=1}^{i-1} j \, \widetilde k_j + i \, \sum_{j=i}^{\widetilde p} 
\widetilde k_j \   . \label{eq:ns}
	}
Notice the relation $2 \, \widetilde N_i - \widetilde N_{i+1} - \widetilde N_{i-1}  = \widetilde k_i$, which encodes the $N_f = 2N_c$ condition for the vanishing of the $\beta$
function in the dual quiver description \cite{Witten:1997sc}.

The puncture contribution to the 't Hooft anomalies of the class $\CS$ SCFTs can be stated in terms of this data as follows:
	\ba{
	(n_v-n_h)^{\text{CFT}}(P_\alpha)&= - \frac{1}{2} \sum_{i=1}^{\widetilde{p}} \widetilde{N}_{i} \widetilde{k}_{i} + \frac{1}{2}\ ,  \label{eq:cft1}\\
	n_v^{\text{CFT}}(P_\alpha)&= - \sum_{i=1}^{\widetilde{p}} \left( N^2-\widetilde{N}_{i}^2 \right)  - \frac{1}{2} N^2  + \frac{1}{2}\ ,\label{eq:cft2}\\
	{k}^{\text{CFT}}_{{SU}(\widetilde{k}_{i})}  &= 2 \widetilde{N}_{i}\ .  \phantom{\frac 12 }  \label{eq:cft3}
	}
The last equation is the mixed flavor-R-symmetry contribution due to a factor $SU(\widetilde{k}_{i})$ of the flavor group.  These contributions were computed explicitly for the $A_n$ case in \cite{Gaiotto:2009gz,Chacaltana:2010ks}, with the general ADE formula derived in \cite{Chacaltana:2012zy}\footnote{Another common notation uses the {\it pole structure}, a set of $N$ integers $p_i$ defined by sequentially numbering each of the $N$ boxes in the Young diagram, starting with 1 in the upper left corner and increasing from left to right across a row such that $p_i = i - $(height of $i$'th box) \cite{Gaiotto:2009we}. These are related to the $\widetilde{N}_{i}$ as 
	\ba{
	\sum_{i=1}^N (2i-1)p_i = \frac{1}{6} \left( 4 N^3 - 3 N^2 - N\right) - \sum_{i=1}^{\widetilde{p}} (N^2- \widetilde{N}_{i}^2).
	}}.

It will also be useful to note the following expressions for $n_v$, $n_h$
associated to a free tensor multiplet reduced on a Riemann surface without punctures:
\beq \label{eq:nv_nh_free_tensor}
n_v  ^{\text{free tensor}} = - \frac 12 \chi(\Sigma_{g,0}) \ , \qquad
(n_v - n_h)^{\text{free tensor}}   =  - \frac 12 \chi(\Sigma_{g,0})
\ . 
\eeq
These expressions can be found by dimensional
reduction of the 8-form anomaly polynomial
of a single M5-brane---see appendix \ref{tensor_appendix} for more details.

\subsection{Relating Inflow Data to Young Diagram Data}

The map between the data of the Young diagram and the inflow data is as follows. 
Consider a profile with $p$ monopoles. The monopole located at ${\eta}_a$ on the $\eta$ axis has charge
	\ba{
	{k}_a = {\ell}_a - {\ell}_{a+1}\ ,
	} 
where we used \eqref{P_and_ka} 
to express $k_a$ in terms of $\ell_a$. Let us recast the sum rule \eqref{flux_sum_rule}
in the form
\beq \label{eq:nicer_sum_rule}
N = \sum_{a=1}^p (w_a - w_{a-1}) \, \ell_a \ .
\eeq
We can interpret 
\eqref{eq:nicer_sum_rule} as a partition of $N$
determined by the Young diagram 
\beq
\mathsf Y 
= [(\ell_1)^{w_1} , (\ell_2)^{w_2- w_1}, \dots, (\ell_a)^{w_a - w_{a-1}} , \dots, (\ell_p)^{w_p - w_{p-1}}] \ .
\eeq
We are using a notation in which $\mathsf Y$
is specified by giving the lengths of its rows.
More precisely, we list the \emph{distinct}
lengths $\ell_a$ in decreasing order.
The exponent of $\ell_a$ is the number
of rows with length $\ell_a$.

\begin{figure}
\centering
\includegraphics[width = 11.cm]{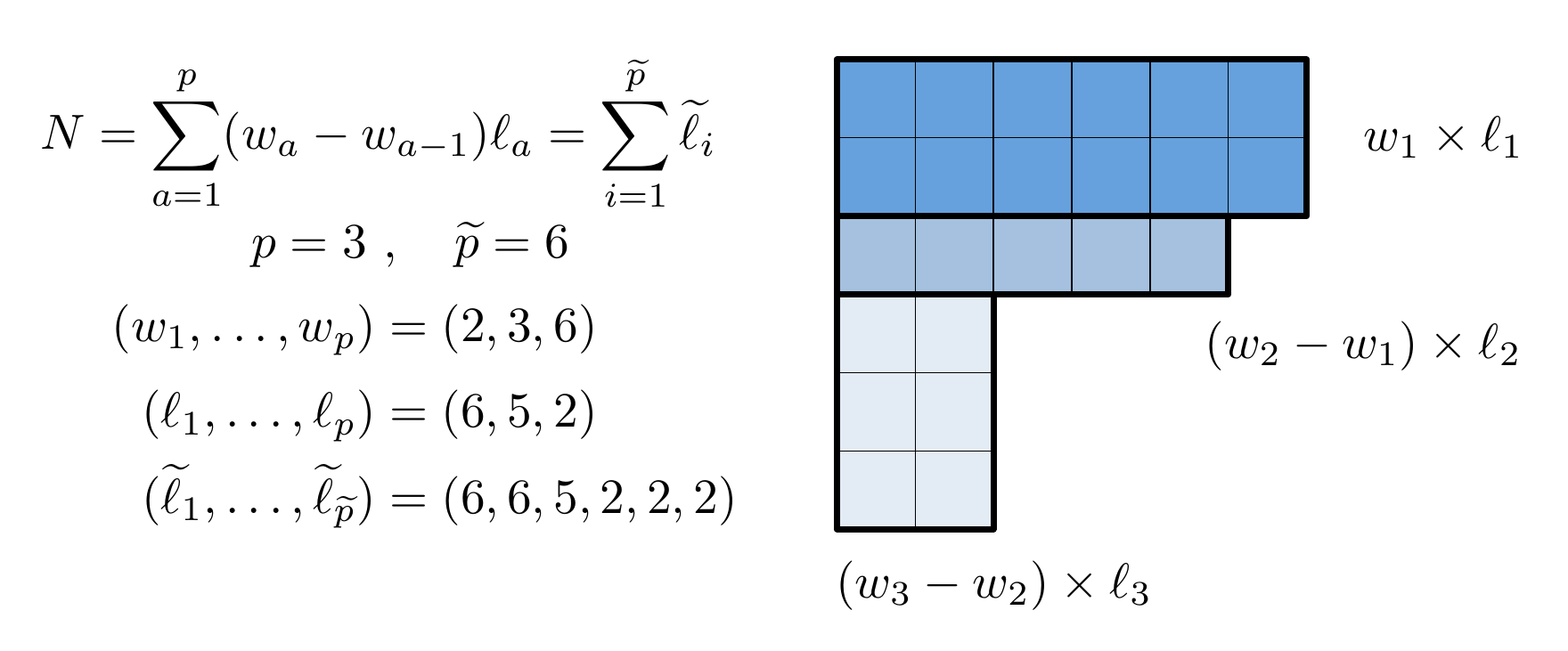}
\caption{
The example of figure \ref{young_from_flux}
is reformulated in terms of $\ell_a$, $\widetilde \ell_i$.
We highlighted the 
decomposition of the Young diagram in  rectangular
blocks
of dimensions $(w_a-w_{a-1}) \times \ell_a$.
}
\label{young_from_flux2}
\end{figure}

The map to the rows $\widetilde{\ell}_i$ of the Young diagram that describes the CFT is
\beq
\widetilde \ell_i = 
\text{value of $L$ along the $\eta$ axis for $\eta \in (i-1, i)$} \ , \qquad 
i = 1 , 2, \dots, w_p \ .
\eeq
Equivalently, we can write
\beq
\widetilde \ell_i = \ell_a \qquad \text{for all $i = w_{a-1}+1 , \dots , w_a$ } \ .
\eeq
Then, the sequence
\beq
( \widetilde \ell_1 \ , \widetilde \ell_2 \ , \dots \ , \widetilde \ell_{\widetilde{p}} ),\qquad \widetilde{p} = w_p\ .
\eeq
is exactly the sequence of lengths of all rows of $\mathsf Y$,
this time listed with repetitions. The total number of rows is equal to the quantity $w_p$. Figure \ref{young_from_flux2} shows the example considered in figure \ref{young_from_flux},
reformulating the partition of $N$ in terms of $\ell_a$ and $\widetilde \ell_i$.

We can identify the monopole charge $k_a$ with the $\widetilde{k}_i$ as
\beq
\widetilde k_i = \left\{ 
\begin{array}{ll}
0 & \qquad \text{if $i \notin \{ w_a\}_{a=1}^p$\ ,} \\[1mm]
k_a & \qquad \text{if $i= w_a$\ .}
\end{array}
\right.
\eeq
When this is nonzero, it corresponds to a location of a monopole, and equals a corresponding $k_{a}$. Therefore we can equivalently rewrite the flavor symmetry \eqref{eq:flavorsymm1}
as 
 	\ba{
	G_F =S\left[  \prod_{a=1}^p  U(k_a)\right]. \label{eq:flavorsymm2}
	}
In this way, the variables that run over  number-of-monopoles and those that run over  number-of-rows are related by taking into account the multiplicity of rows of the same length.

\begin{figure}[t!]
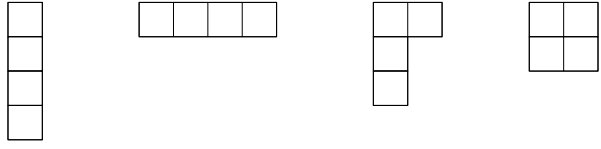

\centering
\yng(1,1,1,1) ~~~~~~~~
 \yng(4) ~~~~~~~~
 \yng(2,1,1)~~~~~~~~
 \yng(2,2) 
\caption{The Young diagrams corresponding to the four discussed examples, drawn with $N=4$ boxes. From left to right, the preserved flavor symmetry is: $\emptyset$, $SU(4)$, $U(1)$, $SU(2)$.\label{fig:young}}
\end{figure}

Before going on, we pause to go through several examples of puncture profiles, mapping the Young diagram data to the inflow data and computing the anomaly contributions of the punctures. We draw the corresponding Young diagrams for the case of $N=4$ in figure \ref{fig:young}.

\paragraph{Example 1: Non-puncture}

The Young diagram data that labels a non-puncture (no flavor symmetry) is:
\ba{\bs{
\text{non-puncture}:\qquad &\widetilde{p} = N,\quad \widetilde{\ell}_{i=1,\dots,N}= 1,\quad (\widetilde{k}_{i=1,\dots,N-1}=0,\ \widetilde{k}_{N}=1),\quad \widetilde{N}_{i}=i\ .
}}
The corresponding inflow data is
\ba{\bs{
\text{non-puncture}:\qquad &p = 1,\quad {\ell}_{1}= 1,\quad {k}_{1}=1,\quad {N}_1=N,\quad {w}_1 = N\ .
}}
For this case, the CFT answers \eqref{eq:cft1}-\eqref{eq:cft3} simplify to
\ba{
(n_v-n_h)^{\rm CFT} (P_{\text{non}} ) & = -\frac{1}{2}(N-1)\ , \\
 n_v^{\rm CFT} (P_{\text{non}} )&= -\frac{1}{6} (4N^3 - N ) + \frac{1}{2}  \ .
}
This has the net effect of shifting  $\chi\to \chi+1$, or in other words, the number of punctures $n$ from $n\to n-1$. This is exactly the behavior of a non-puncture, whose only contribution is ``filling'' a hole on the Riemann surface.

We can compare with the inflow answer. 
Plugging in to \eqref{eq:inflowans}, we obtain
\ba{
(n_v-n_h)^{\rm inflow} (P_{\text{non}} )&=  \frac{1}{2}N \ , \\
n_v^{\rm inflow} (P_{\text{non}} )&= \frac{1}{6} \left( 4N^3 - N \right) \ .
}
Comparing with the bulk inflow answers \eqref{eq:bulk1}, \eqref{eq:bulk2}, we observe agreement up to $\CO(1)$ terms.

\paragraph{Example 2: Maximal puncture}

The puncture that preserves the maximal flavor symmetry of $G_F=SU(N)$ is known as a maximal puncture. In this case the tilde'd variables that denote the Young diagram data are exactly equivalent to the un-tilde'd variables from the geometry since there is both one monopole and one row, and are given by:
	\ba{\bs{
	{SU(N)}:\qquad &p=1,\quad \ell_1=N,\quad k_1=N,\quad N_1=N,\quad w_1=1\ .
	}
	}

The CFT answers are given by \eqref{eq:cft1}-\eqref{eq:cft3}, which for the maximal puncture simplify to
\beq
(n_v-n_h)^{\rm CFT}(P_{\text{max}}) = n_v^{\rm CFT}(P_{\text{max}}) = - \frac 12 (N^2-1)  \ .
\eeq
In comparison, the inflow result is
\beq
(n_v-n_h)^{\rm inflow}(P_{\text{max}}) = n_v^{\rm inflow}(P_{\text{max}}) =  \frac 12 N^2  \ .
\eeq

\paragraph{Example 3: Minimal puncture}

The puncture profile that preserves the minimal flavor symmetry of $U(1)$ corresponds to a Young diagram with
\ba{\bs{
G_F=U(1):\qquad &\widetilde{p} = N-1,\quad (\widetilde{\ell}_1=2,\ \widetilde{\ell}_{i=2,\dots,N-1} = 1)\ ,\\
		& (\widetilde{k}_1=1,\ \widetilde{k}_{2,\dots,N-2}=0,\ \widetilde{k}_{N-1}=1),\quad \widetilde{N}_{i}=i+1\ .
}}
Equivalently, in terms of the inflow data:
	\ba{\bs{
G_F=U(1):\qquad &p = 2,\quad ({\ell}_1=2,\ {\ell}_{2} = 1),\quad ({k}_1=1,\ {k}_{2}=1)\ ,\\
& ({N}_1 = 2,\ {N}_2 = N),\quad ({w}_1=1,\ {w}_2=N-1)\ .
}}
There are $p=2$ monopoles, each with monopole charge 1.

The CFT anomalies  \eqref{eq:cft1}-\eqref{eq:cft3} for the minimal puncture are
	\ba{
	(n_v-n_h)^{\text{CFT}}(P_{\text{min}}) =-\frac{1}{2}(N+1), \quad n_v^{\text{CFT}}(P_{\text{min}})  = - \frac{1}{6} \left( 4N^3-6N^2-N+3\right).
	}
Plugging the inflow data into \eqref{eq:inflowans}, we once again find
\beq
n_v^{\rm inflow}(P_{\text{min}})  + n_v^{\rm CFT}(P_{\text{min}})  = \frac 12 \ , \qquad
n_h^{\rm inflow}(P_{\text{min}})  + n_h^{\rm CFT}(P_{\text{min}})  = 0  \ .
\eeq

\paragraph{Example 4: Rectangular diagram}

For even $N$, we can preserve $SU(N/2)$ via:
\ba{\bs{
G_F=SU(N/2):\qquad &\widetilde{p} = 2,\quad (\widetilde{\ell}_1=N/2,\ \widetilde{\ell}_{2} = N/2)\ ,\\
		& (\widetilde{k}_1=0,\ \widetilde{k}_{2}=N/2),\quad (\widetilde{N}_{1}=N/2,\ \widetilde{N}_2 = N)\ ,
}}
or equivalently, in terms of the inflow data:
	\ba{\bs{
G_F=SU(N/2):\qquad &p = 1,\quad \ell_1=N/2,\quad k_1=N/2,\quad N_1=N,\quad w_1=2\ .
}}
For this case, the CFT puncture anomalies are
\ba{
(n_v-n_h)^{\rm CFT}  (P_{\text{rect}})   = - \frac{1}{4} N^2 + \frac{1}{2}, \qquad 
 n_v^{\rm CFT} (P_{\text{rect}})   = - \frac{5}{4}n^2 + \frac{1}{2} \ .
}
and the inflow puncture anomalies are
\ba{
(n_v-n_h)^{\rm inflow} (P_{\text{rect}})   = \frac{1}{4} N^2,  \qquad
n_v^{\rm inflow} (P_{\text{rect}})  = \frac{5}{4} N^2  \ .
}

\subsection{Matching CFT and Inflow Results}

 Comparing \eqref{eq:bulk1}-\eqref{eq:bulk2} with \eqref{eq:nvnhbulk}-\eqref{eq:nvbulk}, we see that our results for the bulk anomalies can be summarized as  
	\ba{
	n_v^{\rm inflow}(\Sigma_{g,n}) + n_v^{\text{CFT}}(\Sigma_{g,n}) &= \frac{1}{2} \chi(\Sigma_{g,n})\ , \label{eq:resultbulk1}\\
	  n_h^{\rm inflow}(\Sigma_{g,n})+ n_h^{\rm CFT}(\Sigma_{g,n})&=0\ .   \phantom{\frac 12 } \label{eq:resultbulk2}
	  }
Our results for anomalies due to a single puncture on the surface can be summarized as
	\ba{
	n_v^{\rm inflow}(P_\alpha) + n_v^{\text{CFT}}(P_\alpha)  &= \frac{1}{2}\ , \label{eq:result1}\\
	  n_h^{\rm inflow}(P_\alpha)  + n_h^{\rm CFT}(P_\alpha) &=0 
	  \ ,  \phantom{\frac 12 } \label{eq:result2}\\
		k_{SU(k_a)}^{\text{inflow}} + k_{SU(k_a)}^{\text{CFT}}  &= 0\ .  \phantom{\frac 12 } \label{eq:result3}
	}
We prove these relations in appendix \ref{sec:matching_proof} using the mapping discussed in the previous subsection.	
Then, adding up the contribution of all $n$ punctures on the surface 
\`a la
 \eqref{eq:totsn} gives 
	\ba{
	n_v^{\text{inflow}} + n_v^{\rm CFT} = n_v^{\rm free\ tensor}(\Sigma_{g,0}) &= \frac{1}{2}\chi(\Sigma_{g,0})\ , \\
	n_h^{\text{inflow}} + n_h^{\rm CFT} = n_h^{\rm free\ tensor}(\Sigma_{g,0}) &= 0\ .
	 \phantom{\frac 12 } 
	}
We see that the inflow computation exactly cancels the CFT computation, up to the contribution of a single free tensor multiplet over the Riemann surface that  does not see the punctures.



\section{Conclusion and Discussion}
\label{sec_conclusion}

In this work we have considered 4d $\cN = 2$ class $\cS$ theories
obtained from compactification of the 6d $(2,0)$ theory of type $A_{N-1}$
 on a Riemann surface $\Sigma_{g,n}$ with an arbitrary number
 of regular punctures. We have provided a first-principles derivation
 of their 't Hooft anomalies from the corresponding M5-brane setup.
 More precisely, we have shown that anomaly inflow from the M-theory bulk cancels exactly against the CFT anomaly, up to the decoupling modes from a free (2,0) tensor multiplet compactified on the Riemann surface $\Sigma_{g,0}$.

The inflow anomaly polynomial is obtained by integrating the characteristic class
 $\cI_{12}$ over the space $M_6$.  The latter is a smooth geometry supported by non-trivial $G_4$-flux configuration. 
In the absence of punctures $M_6$ is an $S^4$ fibration over the Riemann surface,
but in the presence of punctures it acquires a richer structure. 
 The topology of $M_6$ and the fluxes of
$G_4$ along non-trivial 4-cycles encode all the discrete data of the class 
$\cS$ construction. In particular, the partition of $N$ that labels a regular puncture
is derived from regularity and flux quantization of $G_4$
in the region of $M_6$ near the puncture.

Our inflow analysis has interesting connections to holography.
At large $N$, the holographic dual of an $\cN = 2$
class $\cS$ theory of type $A_{N-1}$ with regular punctures
is given by the Gaiotto-Maldacena solutions of 11d supergravity \cite{Gaiotto:2009gz}.
These solutions are warped products of $AdS_5$ with an internal
6d manifold $M_6^{\rm hol}$, supported by a non-trivial $G_4$-flux
configuration $G_4^{\rm hol}$.
The topology of $M_6^{\rm hol}$ coincides with the topology of $M_6$,
and $G_4^{\rm hol}$ is equivalent in coholomogy to $\overline E_4$,
which is the class $E_4$ with the connections of external spacetime turned off.
We refer the reader to appendix \ref{sec:GMappendix} for more details.
In other words, the classical solution to two-derivative supergravity---which is 
valid at large $N$---provides a local expression for the metric and flux
that is representative of the 
 topological properties of the pair $(M_6,\overline E_4)$
relevant to the inflow procedure---which gives results that are exact in $N$.
This observation is particularly interesting in
light of the fact that, thanks to superconformal symmetry,
the 't Hooft anomaly coefficients are related to the $a$, $c$ central
charges of the CFT. Anomaly inflow thus provides a route to the exact
central charges, which in turn contain non-trivial information
about higher-derivative corrections to the effective action
of the $AdS_5$ supergravity obtained by reducing M-theory
on $M_6^{\rm hol}$.
This circle of ideas admits natural generalizations to other
holographic setups based on 11d supergravity solutions that
describe the near-horizon geometry of a stack of M5-branes,
including $\cN= 1$ constructions such as \cite{Bah:2011vv,Bah:2012dg}.
The interplay between
M5-brane geometric engineering, anomaly inflow, and holography
warrants further investigation.

We believe that the methods of this paper can be generalized to
treat a larger class of punctures. For instance, it would be 
interesting to identify the local geometry and $G_4$-flux
configuration for $\cN = 2$ irregular punctures.
In that case, we expect a more subtle interplay between
bulk and puncture. This intuition is motivated by the fact that,
in setups with irregular punctures, the 4d $U(1)_r$ symmetry
results from a non-trivial mixing of the $S^1_\phi$ circle
with a global $U(1)$ isometry on the Riemann surface
(which is necessarily a sphere) \cite{Gaiotto:2009hg}.

Our strategy can also be  applied to regular
$(p,q)$ punctures in $\cN=1$ class $\cS$ \cite{Bah:2015fwa,Bah:2018gwc}. A $(p,q)$ puncture
preserves locally an $SU(2) \times U(1)$ R-symmetry, 
which is twisted with respect to the 
$SU(2) \times U(1)$ R-symmetry in the bulk of the Riemann surface.
We expect that a regular $(p,q)$ puncture is described by the same
local geometry $X_6$ we constructed for regular $\cN = 2$ punctures.
The gluing prescription of $X_6$ onto $M_6^{\rm bulk}$, however,
is different. The space $X_6$ is a fibration of a 2-sphere $S^2_{\rm punct}$
onto the space $X_4$ spanned by $(\rho, \eta, \chi, \beta)$.
In the usual case, $S^2_{\rm punct}$ is trivially identified
with $S^2_\Omega$ in the bulk.
For a $(p,q)$ puncture,
the angle $\chi$ and the azimuthal angle of $S^2_{\rm punct}$
are rotated in a non-trivial way 
before being identified with the angle $\phi + \beta$
and the azimuthal angle of $S^2_\Omega$ in the bulk, respectively.

We also envision generalizations of our approach to
a broader class of M-theory/string theory constructions.
Our findings reveal that the class $\cI_{12}$ governs the
anomalies of 4d $\cN = 2$ theories obtained from
compactification of the 6d $(2,0)$ theory of type $A_{N-1}$.
We expect that the same class $\cI_{12}$ also governs
the anomalies of many other lower-dimensional theories  obtained
from the same parent theory in six dimensions,
including 4d $\cN = 1$ theories of class $\cS$ type, and 2d SCFTs 
from M5-branes wrapped on four-manifolds.
It is natural to conjecture that
 this framework still holds 
if we replace the 6d $(2,0)$ theory of type $A_{N-1}$
with a different 6d SCFT that 
can be engineered in M-theory using M5-branes.
For example, one may consider the $(2,0)$ theory of type $D_{N}$,
whose anomalies were derived via inflow in \cite{Yi:2001bz}, $(1,0)$ E-string theories, whose anomalies are studied in \cite{Ohmori:2014pca}, or $(1,0)$ SCFTs describing M5-branes 
probing an ALE singularity, with anomalies analyzed in  \cite{Ohmori:2014kda}.
In each case, a single characteristic class
would govern the anomalies of both the parent 6d theory,
and of many lower-dimensional theories obtained
via dimensional reduction of the former. 
One can also consider generalizations of this framework
to other brane constructions in Type IIB/F-theory
and (massive) Type IIA.

Finally, we emphasize that our description of punctures is different 
from and complementary to previous methods that use more field-theoretic tools.  
Indeed, the approach developed here is more readily generalizable in M-theory and string theory,
thus allowing us to
 address a wider class of questions involving anomalies in geometrically engineered field theories.


\section*{Acknowledgments}

We would like to thank Jacques Distler, Thomas Dumitrescu, Simone Giacomelli, Ken Intriligator, David Kaplan, Jared Kaplan, 
Zohar Komargodski,
Craig Lawrie, Mario Martone, Greg Moore, Raffaele Savelli, Sakura Sch{\"a}fer-Nameki, Jaewon Song, Yuji Tachikawa, Alessandro Tomasiello, and Yifan Wang for interesting conversations and correspondence. The work of IB and FB is supported in part by NSF grant PHY-1820784. RM is supported in part by ERC Grant 787320 - QBH Structure. We gratefully acknowledge the Aspen Center for Physics, supported by NSF grant PHY-1607611, for hospitality during part of this work.  


\appendix


\section{Global Angular Forms, Bott-Cattaneo Formula, and $\cI_{12}$} 
\label{angular_appendix}
In this appendix we review some basic properties of global angular forms in
odd-dimensional sphere bundles, following \cite{Freed:1998tg,Harvey:1998bx}.
We also review a useful result of 
Bott and Cattaneo \cite{bott1999integral}. 
Next, we briefly review the derivation of $\cI_{12}$.
Finally, we explore the interplay between the descent formalism
and integrations along the fibers of the sphere bundle.

\subsection{Conventions for Characteristic Classes} 
\label{appendix_classes}

Consider a connection on a $\mathfrak{u}(M)$ bundle with anti-Hermitian field strength $F_{\mathfrak{u}}$. This can be diagonalized by an element of $U(M)$ as
	\ba{
	\frac{i F_{\mathfrak{u}}}{2\pi} = \left( \begin{array}{ccc} \lambda_1 & & \\ & \lambda_2 & \\ & & \ddots \end{array}\right)\ .
	}
For an $\mathfrak{su}(M)$ bundle, $\sum_i\lambda_i=0$. One can define a characteristic polynomial (also called the total Chern class) as 
	\ba{
	c = \text{det}\left(\mathbf{1} + \frac{i F_{\mathfrak{u}}}{2\pi}\right) = 1 + c_1 + c_2  +  \dots
	}
Here the $c_k$ are the  $2k$-form Chern classes, e.g.
	\ba{\bs{
	c_1 = \frac{\tr i F_{\mathfrak{u}}}{2\pi},\quad c_2 = \frac{1}{2(2\pi)^2} \left[ \tr F_{\mathfrak{u}}^2-(\tr F_{\mathfrak{u}})^2  \right]\ .
	}}
Equivalently, we can write
	\ba{
	c_1 = \sum_i \lambda_i\ ,\qquad c_2 = \sum_{i<j}\lambda_i\lambda_j\ .
	}
The Chern character is defined as 
	\ba{
	\text{ch} = \tr_{\mathbf{r}} e^{iF_{\mathfrak{u}}/(2\pi)} = \text{dim}(\mathbf{r}) + c_1 + \frac{1}{2} (c_1^2-2c_2) +  \dots	
	}
Note that in our notation for a $U(1)$ gauge field $A$, $i A_{\mathfrak{u}} = A$, such that $c_1 = \frac{F}{2\pi}$.

The field strength associated to a connection on a real $\mathfrak{so}(2M)$ bundle can be written
	\ba{
	\frac{F_{\mathfrak{so}}}{2\pi} = \left( \begin{array}{ccccc} 0 & \lambda_1 & & & \\ -\lambda_1 & 0 & & &\\ & & 0 & \lambda_2 & \\ & & - \lambda_2 & 0 & \\ & & & & \ddots \end{array}\right)\ .
	}
The Pontryagin classes $p_k$ are $4k$-forms, e.g.
	\ba{
	p_1 = - \frac{1}{2(2\pi)^2} \tr F_{\mathfrak{so}}^2\ ,\qquad p_2 = \frac{1}{8(2\pi)^4} \left[ (\tr F_{\mathfrak{so}}^2)^2-2\tr F_{\mathfrak{so}}^4\right]\ .
	}
These are packaged into a characteristic polynomial as
	\ba{
	p = \text{det}\left(\mathbf{1} + \frac{F_{\mathfrak{so}}}{2\pi}\right)  = 1 + p_1 + p_2 + p_3 + \dots
	}
The Pontryagin classes can be written in terms of the Chern roots $\lambda_i$ as
	\ba{
	p_1 = \sum_j \lambda_j^2,\qquad p_2 = \sum_{i<j} \lambda_i^2\lambda_j^2,\qquad \dots
	}	
Another useful set of identities relates the Pontryagin calsses of a Whitney sum of two vector bundles $E = E_1 \oplus E_2$ to the Pontryagin classes of the constituents, as
	\ba{
	p_1(E) &= p_1(E_1) + p_1(E_2)\ ,\qquad p_2(E) = p_2(E_1) + p_2(E_2) + p_1(E_1)p_1(E_2)\ .\label{eq:psum}
	}

\subsection{Global Angular Forms} 
\label{appendix_global_ang_forms}

Let $\cE$ be a real vector bundle 
of odd rank $2m+1$ over a base space $B$.
The fiber of $\cE$ over a point $p  \in B$ is a copy of
$\mathbb R^{2m+1}$, parametrized by Cartesian coordinates $y^A$,
$A = 1, \dots, 2m+1$,
and equipped with the fiber metric $\delta_{AB}$.
Let $S(\cE)$ be the associated sphere bundle.
For our purposes, the latter is most conveniently
thought of as the bundle over $B$ whose fiber over a point $p$
is the unit $S^{2m}$ sphere inside the $\mathbb R^{2m+1}$ fiber of $\cE$ over $p$.
The sphere $S^{2m}$ is defined by the relation
\beq \label{sphere_eq}
 \hat y_A \, \hat y^A = 1 \ ,
\eeq
where indices $A$, $B$, etc.~are raised and lowered with $\delta_{AB}$.
We have included a hat as a reminder that 
the coordinates $\hat y^A$ are henceforth understood
to obey the  constraint  \eqref{sphere_eq}.

Working with these local coordinates,
the non-triviality of the $S(\cE)$ fibration is encoded in the
covariant differentials
\beq
D\hat y^A = d\hat y^A - \Theta^{AB} \, \hat y_B \ ,
\eeq
where $\Theta^{AB}$ are the components of 
a $\mathfrak{so}(2m+1)$ connection over the base space $B$.
Notice that the volume form on the fiber sphere is
\beq
{\rm vol}_{S^{2m}} = \frac{1}{2 \, m! \, (4\pi)^m} \, \epsilon_{A_1 \dots A_{2m+1}}  \, y^{A_1} \, d\hat y^{A_2} \wedge \dots \wedge d\hat y^{A_{2m+1}}   \ ,
\eeq
where we selected the prefactor in such a way
that ${\rm vol}_{S^{2m}}$ integrates to $1$.
The form ${\rm vol}_{S^{2m}}$ is closed but it is not invariant under 
the action of the $SO(2m+1)$ structure group of the fibration.
In this language, the global angular form is a $2m$-form
$E_{2m}$ which is the unique closed and gauge-invariant
improvement of ${\rm vol}_{S^{2m}}$. The class $E_{2m}$ 
can be written as
\beq
E_{2m} =  \frac{1}{2 \, m! \, (4\pi)^m}  \, \epsilon_{A_1 \dots A_{2m+1}}  \, y^{A_1} \, D\hat y^{A_2} \wedge \dots \wedge D\hat y^{A_{2m+1}}
  + P_{2m}(\hat y , D\hat y, F) \ ,
\eeq
where the corrective term $P_{2m}(\hat y , D\hat y, F)$ is a polynomial
in $\hat y^A$, $D\hat y^A$, and $F^{AB}$, which are the components
of the field strength of the $\mathfrak{so}(2m+1)$ connection,
\beq
F^{AB} = d \Theta^{AB} - \Theta^{AC} \wedge \Theta_C{}^B \ .
\eeq
The corrective term $P_{2m}(\hat y , D\hat y, F)$ is given explicitly for any $m$ in \cite{Harvey:1998bx}. 
Let us record here only the full expressions for $m=1$ and $m=2$,
\begin{align} \label{E2andE4}
E_2 &\equiv e_2^\Omega = \frac{1}{8 \pi}  \Big[ \epsilon_{A_1 A_2 A_3} \, D\hat y^{A_1} \, D\hat y^{A_2}  
\hat y^{A_3} - \epsilon_{A_1 A_2 A_3} \, F^{A_1A_2} \, \hat y^{A_3} \Big] \ ,  
\nn \\
E_4 &= \frac{1}{64 \pi^2}  \Big[ \epsilon_{A_1 \dots A_5} \,   D\hat y^{A_1}  \,
D\hat y^{A_2} \, D\hat y^{A_3} \, D\hat y^{A_4} \, \hat y^{A_5}
- 2 \, \epsilon_{A_1 \dots A_5} \, F^{A_1 A_2} \, D\hat y^{A_3} \,
D\hat y^{A_4} \, \hat y^{A_5} \nn \\
&
+ \epsilon_{A_1 \dots A_5} \, F^{A_1A_2} \, F^{A_3 A_4} \, \hat y^{A_5} \bigg] \ .
\end{align}
Clearly, the range of $A$ indices in the first relation is from 1 to 3,
and in the second is from 1 to 5.
For brevity, we have suppressed wedge products.
In the first relation we have made contact with the notation $e_2^\Omega$ used 
in the main text for the global angular form for $SO(3)$.
Let us stress that in writing down the above formula for $E_4$
we have made the assumption of an unbroken
structure group $SO(5)$. In the main text, the structure group
is reduced, and hence $E_4$ takes a different form,
see \eqref{general_E4}.

\subsection{Bott-Cattaneo Formula}
\label{appendix_bott_cattaneo}

The Bott-Cattaneo formula \cite{bott1999integral}
gives the integral of any power of the global angular form 
$E_{2m}$ along the $S^{2m}$ fiber directions.
The formula reads
\begin{align} \label{BottCattaneoFormula}
\int_{S^{2m}} (E_{2m})^{2s+2} = 0 \ , \qquad
\int_{S^{2m}} (E_{2m})^{2s+1} = 2^{-2s} \, \Big[p_m(\cE) \Big]^s \ , \qquad
s = 0,1, 2, \dots
\end{align}
The symbol $p_m(\cE)$ denotes the standard Pontryagin classes
of the vector bundle $\cE$.
Let us stress that we are using conventions in which
$E_{2m}$ integrates to $1$ on the $S^{2m}$ fibers.
(In the mathematics literature, $E_{2m}$ usually integrates
to $2$.)

\subsection{Derivation of $\cI_{12}$}
\label{sec_I12_derivation}
In this subsection we summarize the arguments of \cite{Freed:1998tg,Harvey:1998bx}
leading to the introduction of the characteristic class $\cI_{12}$.
Our starting point is the Bianchi identity \eqref{dG4_generic}, repeated here for
convenience,
\beq
\frac{dG_4}{2\pi} = df \wedge E_4 \ .
\eeq
Since the RHS is non-zero, the standard relation $G_4 = dC_3$ is modified to
\beq
\frac{G_4}{2\pi} = \frac{dC_3}{2\pi} - df \wedge E_3^{(0)} \ , \qquad
dE_3^{(0)} = E_4\ .
\eeq
Let us stress that $E_3^{(0)}$ is \emph{not} gauge-invariant
under $SO(5)$ transformations. Indeed, descent gives
\beq
\delta E_3^{(0)} = dE_2^{(1)} \ .
\eeq
Since $G_4$ must be gauge-invariant under $SO(5)$ transformations, $C_3$ must acquire an anomalous
gauge variation under $SO(5)$ transformations, 
\beq
\frac{\delta C_3}{2\pi} = -  df \wedge E_2^{(1)}  \ .
\eeq
The above relation suggests an improvement of  $C_3$,
denoted $\widetilde C_3$, whose anomalous gauge variation
is a total derivative,
\beq
\frac{\widetilde C_3}{2\pi} = \frac{  C_3}{2\pi} - f \, E_3^{(0)} \ , \qquad
\frac{ \delta \widetilde C_3}{2\pi}  = d \Big[
- f \, E_2^{(1)}
\Big] \ .
\eeq
Given the gauge transformation law of $\widetilde C_3$,
the following quantity is gauge invariant,
\beq \label{tildeG4}
\frac{\widetilde G_4}{2\pi} = \frac{d \widetilde C_3}{2\pi}
= \frac{dC_3}{2\pi} - df \wedge E_3^{(0)}
- f \, E_4 \ .
\eeq

Recall that, upon regularizing the delta-function singularity
in the Bianchi identity for $G_4$, we excise a small tubular
neighborhood $B_\epsilon$ of radius $\epsilon$ of the M5-brane stack.
The 11d M-theory effective action is now formulated on a spacetime
with a boundary $S^4 \hookrightarrow X_{10} \rightarrow W_6$. The only relevant terms are the topological
couplings $C_3 G_4 G_4$ and $C_3 I_8$,
where $I_8$ is the characteristic class \eqref{I8def}.
More precisely,
\beq
\frac{S_M}{2\pi} \supset \int_{M_{11} \setminus B_\epsilon} \bigg[
- \frac 16 \, \frac{\widetilde C_3 \, \widetilde G_4 \, \widetilde G_4}{(2\pi)^3} 
- \frac{ \widetilde C_3}{2\pi} \, I_8  \bigg] \ ,
\eeq
where we suppressed wedge products for brevity.
Notice that we have replaced $C_3$ with $\widetilde C_3$,
and accordingly $G_4$ with $\widetilde G_4$.
The gauge variation of the effective action is
\beq
\frac{\delta S_M}{2\pi} = \int_{M_{11} \setminus B_\epsilon} \bigg[
- \frac 16 \, \frac{
\delta \widetilde C_3 \, \widetilde G_4 \, \widetilde G_4
}{2\pi}
- \frac{ \delta \widetilde C_3}{2 \pi} \, I_8  \bigg]
= \int_{M_{11} \setminus B_\epsilon} d\Big[ - f \, E_2^{(1)} \Big] \bigg[
- \frac 16 \,  \frac{ \widetilde G_4 \, \widetilde G_4 }{2\pi}
-  I_8  \bigg] \ .
\eeq
We may now collect a total derivative, and recall
$\partial(M_{11} \setminus B_\epsilon)  = X_{10}$,
see \eqref{X10boundary}. 
The boundary is located at fixed radial coordinate
$r=\epsilon$, and  therefore we can set $f = -1$.
We thus arrive~at
\beq
\frac{\delta S_M}{2\pi} 
= \int_{X_{10}} E_2^{(1)}  \bigg[ 
- \frac 16 \,  \frac{ \widetilde G_4 \, \widetilde G_4 }{2\pi}
-  I_8  \bigg] \ .
\eeq
Since $X_{10}$ sits at $r = \epsilon$, 
we can set $f=-1$ and $df=0$ in \eqref{tildeG4}.
The term $dC_3/(2\pi)$ in
\eqref{tildeG4} is topologically trivial and is neglected. 
We  conclude that
\beq \label{variation_and_I10}
\frac{\delta S_M}{2\pi} 
= \int_{X_{10}} E_2^{(1)}  \bigg[ 
- \frac 16 \,  E_4 E_4
-  I_8  \bigg] 
\equiv \int_{X_{10} }  \cI_{10}^{(1)} \ .
\eeq
Since both $E_4 \, E_4$ and $I_8$ are closed and gauge-invariant
8-forms, the 10-form $\cI_{10}^{(1)}$ satisfies the descent equations
\beq
d \cI_{10}^{(1)} = \delta \cI_{11}^{(0)} \ , \qquad
d\cI_{11}^{(0)} = \cI_{12} = - \frac 16 \, E_4 \, E_4 \, E_4 - E_4 \, I_8 \ . 
\eeq

\subsection{Descent Formalism and Integration Along $S^4$ Fibers}
\label{sec_descent_commutes}
In order to connect \eqref{variation_and_I10}
to the anomaly polynomial of the theory living on the M5-brane stack,
we have to perform the integral 
 over $X_{10}$   in two steps:
we first integrate along the $S^4$ fiber, and then
integrate along the worldvolume $W_6$. To carry out this program,
we need to choose a representative of $\cI_{10}^{(1)}$
that is globally defined on the $S^4$ fibers
(but not necessarily on $W_6$).
Let us write $E_4$ as
\beq
E_4 = {\rm vol}_{S^4} + Z_4 \ , \qquad Z_4 = dZ_3^{(0)} \ ,
\eeq
where $ {\rm vol}_{S^4}$ is the \emph{ungauged}
volume form on $S^4$ (normalized to 1)
and $Z_4$ collects all the terms proportional to the connection $\Theta$
or its field strength $F$. Notice that $Z_4$ is closed, but not gauge-invariant.
We can write $Z_4 = dZ_3^{(0)}$, where $Z_3^{(0)}$
is globally defined on the $S^4$ fibers, is not gauge invariant,
and vanishes if the connection $\Theta$ is set to zero.
We can perform descent of the class $(E_4)^3$
using quantities that are globally defined on $S^4$.
Indeed,  one has
\beq \label{clever_descent}
\Big[ (E_4)^3 \Big]^{(0)} = E_4^2 \, (E_4 + {\rm vol}_{S^4}) \, Z_3^{(0)} \ , \qquad
\Big[ (E_4)^3 \Big]^{(1)}  = E_4 \, (E_4 + 2 \, {\rm vol}_{S^4}) \, Z_3^{(0)} \, \delta Z_3^{(0)}
 \ .
\eeq
To check the
above descent relations, it is useful to recall that
\beq
{\rm vol}_{S^4}^2 = 0  \ , \qquad
0 = \delta E_4 = \delta{\rm vol}_{S^4} + d\delta Z_3^{(0)} \ , \qquad
{\rm vol}_{S^4} \, \delta {\rm vol}_{S^4} = 0 \ . 
\eeq
Thanks to the fact that all quantities in \eqref{clever_descent}
are globally defined on $S^4$, we can make sense of the following
formal manipulations. First of all, let us write the descent relations for
$(E_4)^3$ by splitting the differential into the internal $S^4$ part and the external
part,
\beq \label{argument_before}
(E_4^3)^3 = (d_{\rm ext} + d_{\rm int})\Big[ (E_4)^3 \Big]^{(0)} \ , \qquad
\delta \Big[ (E_4)^3 \Big]^{(0)} =  (d_{\rm ext} + d_{\rm int})\Big[ (E_4)^3 \Big]^{(1)} \ .
\eeq
Let us integrate both these relations on $S^4$. Since 
$\Big[ (E_4)^3 \Big]^{(0)}$ and $\Big[ (E_4)^3 \Big]^{(1)}$
are globally defined on $S^4$, we can invoke Stokes' theorem,
and drop the $d_{\rm int}$ terms. We thus arrive at
\beq \label{argument_after}
\int_{S^4}(E_4^3)^3 =  d_{\rm ext}  \int_{S^4 }\Big[ (E_4)^3 \Big]^{(0)} \ , \qquad
\delta \int_{S_4} \Big[ (E_4)^3 \Big]^{(0)} =  
d_{\rm ext}  \int_{S^4} \Big[ (E_4)^3 \Big]^{(1)} \ .
\eeq
The above relations establish that descent and $S^4$ integration commute.

By a similar token, we perform descent on the $E_4 I_8$ term as
\beq
(E_4 \, I_8)^{(0)} = E_4 \, I_7^{(0)} \ , \qquad
(E_4 \, I_8)^{(1)} = E_4 \, I_6^{(1)} \ .
\eeq
Since the $E_4$ factor is left intact, these quantities are globally defined
on $S^4$, and we can repeat the above argument to
show that descent and $S^4$ integration commute.

In this paper we also consider setups 
of the form $M_6 \hookrightarrow X_{10} \rightarrow W_4$.
The space $M_6$ is a smooth compact manifold.
The gauge variation that enters the descent relations
has a gauge parameter that depends on $W_4$ only.
In this case, the main observation is that it is possible to
find a representative of 
$\cI_{10}^{(1)}$ that is globally defined on $M_6$.
Once such a representative is found,
we can repeat the argument from \eqref{argument_before}
to \eqref{argument_after}, with $S^4$ replaced by $M_6$,
and conclude that descent and integration over $M_6$ commute.

 \subsection{Computation of $I_6^{\rm inf}(\Sigma_{g,n})$}
\label{appendix_I6bulk}
In this subsection we compute $I_6^{\rm inf}(\Sigma_{g,n} )= \int_{M_6^{\rm bulk}} \cI_{12}$.
Let us first consider the term $(E_4)^3$ in $\cI_{12}$.
We can use the Bott-Cattaneo formula \eqref{BottCattaneoFormula} 
to integrate over $S^2_\Omega$,
\beq
\int_{M_6^{\rm bulk}}  (E_4)^3= \frac 14 p_1(N_{SO(3)}) \int_{[\mu] \times S^1_\phi \times \Sigma_{g,n}} (\cE_2)^3  \ ,
\eeq
where we have denoted schematically the residual four directions
of integration. The relevant terms in $(\cE_2)^3$ are
\beq
(\cE_2)^3 \supset  N^3  \, (2\pi)^{-3}\, d(\gamma^3) \, D\phi \, \cF^2
\supset  2 \, N^3  \, (2\pi)^{-3}\, d(\gamma^3) \, D\phi \, F_\phi \, F_\Sigma \ .
\eeq
This is readily integrated recalling $\gamma(0) = 0$, $\gamma(1) =1$, 
$\int_{\Sigma_{g,n}} F_\Sigma = - 2\pi \chi(\Sigma_{g,n})$. We thus get
\beq \label{partial_bulk1}
\int_{M_6^{\rm bulk}}  (E_4)^3= - \frac 12 \, N \, \chi(\Sigma_{g,n}) \,
 p_1(N_{SO(3)})  \, \frac{F_\phi}{2 \pi}   \ .
\eeq
We can now turn to the term $E_4 \, X_8$ in $\cI_{12}$.
The integral over the $S^4$ fibers of $M_6^{\rm bulk}$
is saturated by $E_4 \supset N \, {\rm vol}_{S^4}$, 
\beq
\int_{M_6^{\rm bulk}}   E_4 \, X_8 = N \, \int_{\Sigma_{g,n}} X_8 \ .
\eeq
To evaluate the class $X_8$ we need the decomposition
of the 11d tangent bundle restricted to the brane worldvolume,
\beq
TM_{11}|_{W_6} = TW_4 \oplus T \Sigma_{g,n} \oplus N_{SO(2)} \oplus N_{SO(3)} \ .
\eeq
Recall that the Chern root of $\Sigma_{g,n}$ is $\hat t$,
the Chern root of $N_{SO(2)}$ is $\hat n = - \hat t +\hat n^{\rm 4d}$.
We can now use repeatedly the standard relations
for the Pontryagin classes of a sum of bundles, given in \eqref{eq:psum}.
We obtain
\beq
X_8= \frac{1}{48} \, \hat t \, \hat n^{\rm 4d} \Big[ p_1(TW_4) 
+ p_1(N_{SO(3)}) - (\hat n^{\rm 4d})^2 \Big] + \dots \ ,
\eeq
where we have only included the terms with one $\hat t$ factor.
We then have
\beq \label{partial_bulk2}
\int_{M_6^{\rm bulk}}   E_4 \, X_8 = \frac{1}{48} \,  N \, \chi(\Sigma_{g,n})  \,
\hat n^{\rm 4d} \Big[ p_1(TW_4) 
+ p_1(N_{SO(3)}) - (\hat n^{\rm 4d})^2 \Big]  \ .
\eeq
Using the definition of $\cI_{12}$, \eqref{I12_def},
and the partial results \eqref{partial_bulk1}, \eqref{partial_bulk2}, we recover the
expression  
\eqref{I6inf_bulk_result} for $I_6^{\rm inf}(\Sigma_{g,n})$
given in the main text.


\section{Evaluation of the Integral for the $(E_4)^3$ Term}
\label{integral_for_E4cube}
\begin{figure}
\centering
\includegraphics[width = 4.7cm]{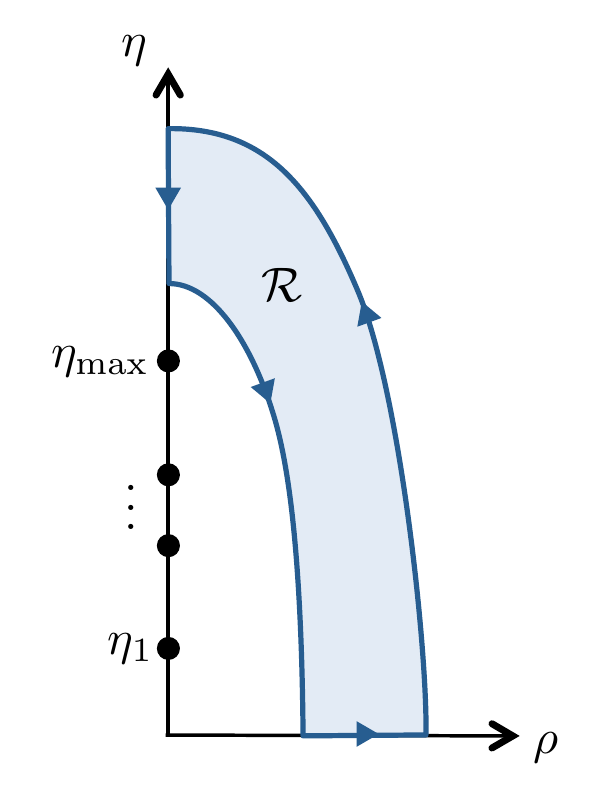}
\caption{
The region $\cR$ in the $(\rho, \eta)$ plane. 
We also depict the boundary $\partial \cR$ with its
positive (counterclockwise) orientation.
}
\label{appendix_fig}
\end{figure}

In the computation of the anomaly inflow from the cubic term $(E_4)^3$
in the puncture geometry
we encounter the following 2-form in the $(\rho, \eta)$ plane,
\beq \label{my_omega2}
\omega_2 = -  3 \, d\Big[ (Y + L  \, W)^2 \Big] \wedge dW \ .
\eeq
Let us integrate $\omega_2$ in the shaded region $\cR$
in the $(\rho, \eta)$ plane  depicted in  figure \ref{appendix_fig},
\beq
\int_\cR \omega_2 =  \int_{\partial \cR} \omega_1 \ , \qquad
\omega_1 :=   3 \, W \, d\Big[ (Y + L  \, W)^2 \Big]
\ .
\eeq
The boundary $\partial \cR$ consists of two arcs and two segments.
The form $\omega_1$ evaluated on the horizontal segment 
gives zero, because $W=0$ for $\eta = 0$.
Moreover, $\omega_1$ is zero on the vertical segment.
This can be seen noticing that, at $\rho=0$ for $\eta > \eta_{\rm max}$,
we have $Y + L  \, W=N$ constant.
It follows that the integral receives contributions from the two arcs only.\footnote{Instead
of $\omega_1$, one
may consider 
\beq
\omega_2  = d\widetilde \omega_1 \ , \qquad \widetilde \omega_1 = - 3 \, (Y + L \, W)^2 \, dW   \ .
\eeq
In this case, however, we get a non-zero contribution from the vertical segment,
since, taking the limit $\rho \rightarrow 0$ with fixed $\eta > \eta_{\rm max}$,
one finds
\beq
\widetilde \omega_1 \approx  - 3 \, N^2 \, dW \ .
\eeq
A contribution from the vertical segment of $\partial \cR$ spoils
the separation between bulk and puncture contributions to the integral.
Therefore, $\widetilde \omega_1$ is not a viable choice,
and we must use $\omega_1$.
}
Notice that the contribution from the large arc does not go to zero
as we increase the size of the arc.
The interpretation is the following. The large arc represents the
bulk contribution to $(E_4)^3$, which is already accounted for
separately in our discussion.
The small arc is identified with the contribution to $(E_4)^3$
localized at the puncture.

Crucially, the integral of $\omega_1$ along the small arc
tends to a finite value as the arc gets closer to the
interval $(0, \eta_{\rm max})$ along the $\eta$ axis. 
The limiting value of $\int \omega_1$ on the small arc
is extracted as follows.

Let us split the interval $(0, \eta_{\rm max})$   into the sub-intervals
$(\eta_{a-1}, \eta_a)$. Recall that $L$ and $Y$ are constant
in each $(\eta_{a-1}, \eta_a)$ interval.
As a result,
\beq
  \int_{(\eta_{a-1},\eta_{a})} 3 \, W \, d \Big[ (Y + W \, L)^2 \Big]
=   \Big[
L \, W^2 (   2 \, L \, W + 3 Y) 
\Big]_{\eta= \eta_{a-1}}^{\eta = \eta_a} \ .
\eeq
Recall that, as $\eta \rightarrow \eta_a$ from below, $Y$ is constant,
$L = \ell_a$, $W \rightarrow  w_a$, and $Y + L \, W \rightarrow N_a$.
It follows that the constant value of $Y$ in the 
$(\eta_{a-1}, \eta_a)$ interval must be $Y = N_a - w_a \, \ell_a$.
As a result,
\beq
  \Big[
L \, W^2 ( 2 \, L \, W + 3 Y) 
\Big]_{\eta= \eta_{a-1}}^{\eta = \eta_a}
=    2 \, \ell_a^2 \, (w_a^3 - w_{a-1}^3)
+ 3 \, \ell_a \, (N_a - w_a \, \ell_a) \, (w_a^2 - w_{a-1}^2) \ .
\eeq
We conclude that
\beq
\int \omega_1 = - \sum_{a=1}^p \bigg[
  2 \, \ell_a^2 \, (w_a^3 - w_{a-1}^3)
+ 3 \, \ell_a \, (N_a - w_a \, \ell_a) \, (w_a^2 - w_{a-1}^2) 
\bigg] \ .
\eeq
Notice that an additional minus sign originates from the fact that $\partial \cR$
is positively oriented if considered counterclockwise,
which induces the negative orientation along the $\eta$ axis.

One might wonder if the integral on the small arc
can pick up contributions localized at the monopoles.
Let us introduce coordinates $R_a$, $\tau_a$ via
\beq
\eta = \eta_a + R_a \, \tau_a \ , \qquad
\rho = R_a \, \sqrt{1- \tau_a^2} \ , \qquad
-1 \le \tau_a \le 1 \ .
\eeq
Restricted on $R_a = \text{const}$, the form $\omega_1$ reads
\beq
\omega_1 = - 3 \, W \, \partial_{\tau_a} \Big[ (Y + L \, W)^2 \Big] \, d\tau_a
= \partial_{\tau_a} \Big[ - 3 \, W \,   (Y + L \, W)^2 \Big] d\tau_a
+ 3 \, (Y + L \, W)^2 \, \partial_{\tau_a} W \, d\tau_a \ .
\eeq
This quantity has to be integrated from $\tau_a =-1$
to $\tau_a = 1$. The first term 
gives clearly
\beq
\Big[ - 3 \, W \,   (Y + L \, W)^2 \Big]_{\tau_a  = -1}^{\tau_a = 1} \ ,
\eeq
and this quantity goes to zero as $R_a \rightarrow 0$,
because both $W$ and $Y + L \, W$ are continuous
across $\eta = \eta_a$ (even though their derivatives
have a discontinuity).
In order to analyze the second term in $\omega_1$,
we notice that
\beq
W = w_a + R_a (a_1 + a_2 \, \tau) + \cO(R_a^2) \ ,
\eeq
where $a_{1,2}$ are constant depending on 
monopole data. 
The quantity $(Y + L \,W)^2$ has a finite 
value as $R_a \rightarrow 0$,
\beq
(Y + L \, W)^2 = N_a^2 + \cO(R_a) \ .
\eeq
At leading order in $R_a$ we thus have
\beq
 3 \, (Y + L \, W)^2 \, \partial_{\tau_a} W \, d\tau_a
= - 3 \, N_a^2 \, R_a \, a_2 \, d\tau_a \ .
\eeq
This quantity has a non-zero integral on $[-1,1]$,
but it is suppressed by the explicit factor of $R_a$.
In summary, we do not expect any 
localized contributions to $\int \omega_1$
from monopole sources.


\section{Free Tensor Anomaly Polynomial}
\label{tensor_appendix}
In this appendix we dimensionally reduce the anomaly
polynomial of a single M5-brane on a Riemann surface $\Sigma_{g,0}$ with no punctures.
The starting point is the 6d anomaly polynomial
\beq
I_8 = \frac{1}{48} \bigg[ p_2(NW_6) - p_2(TW_6) +
\frac 14  \Big(  p_1(NW_6) - p_1(TW_6) \Big)^2 \bigg] \ .
\eeq
The bundles $TW_6$, $NW_6$ decompose as
\beq
TW_6 = TW_4 \oplus T\Sigma_{g,0} \ , \qquad
NW_6 = N_{SO(2)} \oplus N_{SO(3)} \ .
\eeq
As usual, the Chern root of $T\Sigma_{g,0}$ is $\hat t$,
and the Chern root of $N_{SO(2)}$ is $\hat n = - \hat t + \hat n^{\rm 4d}$.
Making use of \eqref{eq:psum},
and collecting all terms linear in $\hat t$,
we arrive at
\beq
I_8 \supset \frac {\hat t}{48} \bigg[
 - 2 \, \hat n^{\rm 4d} \, p_1(N_{SO(3)})
 -  \, \hat n^{\rm 4d}  \, 
 \bigg(
p_1(N_{SO(3)})
- p_1(TW_4) + (\hat n^{\rm 4d})^2 
\bigg)
\bigg] \ .
\eeq
Upon integration over $\Sigma_{g,0}$, the factor $\hat t$ is replaced with
$\chi(\Sigma_{g,0})$.
Making use of the identifications \eqref{put_4d_Cherns},
we get the final result,
\beq
I_6^\text{free tensor} = \frac 1 2  \chi(\Sigma_{g,0}) \, c_1^r \, c_2^R
  - \frac 1 2  \chi(\Sigma_{g,0}) \, \bigg[
\frac 13 \, (c_1^r)^3
- \frac{1}{12} \, c_1^r \, p_1(TW_4)
\bigg] \ .
\eeq
In the parametrization 
\eqref{eq:CFT_general_anomaly} given in terms of $n_{v,h}$,
we have equivalently
\beq
n_v^{\text{free tensor}}= - \frac 12 \, \chi (\Sigma_{g,0}) \ , \qquad
n_h^{\text{free tensor}}= 0 \ .
\eeq



\section{Review of Gaiotto-Maldacena Solutions} \label{sec:GMappendix}
In this appendix we briefly review the Gaiotto-Maldacena (GM)
solutions \cite{Gaiotto:2009gz}, and we clarify their connection with the
inflow setup in the presence of punctures discussed in the main text.

The most general solution to 11d supergravity preserving 4d $\cN = 2$
superconformal symmetry takes the form
\begin{align} \label{GM_Toda_frame}
ds^2_{11} & = \kappa^{2/3} \, e^{2 \widetilde \lambda}  \, \bigg[ 4 \, ds^2_{AdS_5}
+ y^2 \, e^{-6 \widetilde \lambda} \, ds^2_{S^2} 
+ \frac{4 \, (d\phi + v)^2 }{1 - y \, \partial_y D} 
- \frac{\partial_y D}{y} \Big(
dy^2 + e^D \, (dx_1^2 + dx_2^2)
\Big)
 \bigg] \ , \nn \\
G_4^{\rm GM} & = 2\, \kappa \, {\rm vol}_{S^2} \wedge \bigg[
(d\phi + v) \, d(y^3 \, e^{- 6 \widetilde \lambda})
+ y  \, (1 - y^2 \, e^{-6 \widetilde \lambda}) \, dv
- \frac 12 \, \partial_y e^D \, dx_1 \wedge dx_2 \bigg] \ ,
\end{align}
where $\kappa$ is a normalization constant,
$ds^2_{AdS_5}$ is the metric on the unit-radius $AdS_5$,
$ds^2_{S^2}$ is the metric on the round unit-radius $S^2$,
${\rm vol}_{S^2}$ is the corresponding volume form,
the angle $\phi$ has periodicity $2\pi$,
and the function $\widetilde \lambda$ and the 1-form $v$
are determined in terms of the function $D = D(y, x_1, x_2)$ via
\beq
e^{-6 \widetilde \lambda} = - \frac{\partial_y D}{y \,(1 - y \, \partial_y D)} \ , \qquad
v = \frac 12 \, (\partial_{x_2} D \, dx_1 - \partial_{x_1} D \, dx_2) \ .
\eeq
The function $D$ is required to satisfy the Toda equation
\beq \label{Toda_eq}
\Big( \partial^2_{x_1} + \partial^2_{x_2} \Big) D  + \partial_y^2 e^D = 0 \ .
\eeq
In the class $\cS$ context, the metric in \eqref{GM_Toda_frame}
is interpreted as the near-horizon geometry of a stack of M5-branes
wrapping a compact Riemann surface, parametrized by local coordinates
$x_1$, $x_2$. In the case of a Riemann surface with no punctures and genus $g>1$,
the relevant solution to the Toda
equation \eqref{Toda_eq} is
\beq \label{nonpuncture_D}
e^D = \frac{4 \, (N^2 -y^2)}{(1 - x_1^2 - x_2^2)^2} \ .
\eeq
With this choice of $D$, the directions $x_1$, $x_2$ parametrize
a hyperbolic space of constant negative curvature.
The Riemann surface  
is realized as usual by taking a discrete quotient of this hyperbolic space.
The coordinate $y$ parametrizes the interval $[0,N]$,
with the round $S^2$ shrinking at $y = 0$, and the $\phi$ circle $S^1_\phi$ shrinking
at $y = N$. It follows that $y$, $S^1_\phi$, $S^2$ parametrize the $S^4$
surrounding the M5-brane stack. 
From the function $D$ in \eqref{nonpuncture_D}, we compute
\beq \label{Nflux_GM}
\frac{G_4^{\rm GM}}{2\pi}   = \frac{\kappa \, {\rm vol}_{S^2}}{\pi}  \, 
\bigg[ (d\phi +v) \, d \frac{2 \, y^3}{y^2 + N^2}
- \frac{2 \, y^3}{y^2 + N^2} \, dv \bigg]
\ , \qquad 
\int_{S^4} \frac{G_4^{\rm GM}}{2\pi} = 8  \pi  \kappa \, N \ .
\eeq
In order to identify the quantity $N$ with the integer counting the number of M5-branes
in the stack, we need to choose 
 $\kappa = (8\pi)^{-1}$, in accordance with 
our conventions for $G_4$-flux quantization (which are different
from the conventions of \cite{Gaiotto:2009gz}).

In the inflow setup, the $S^4$ surrounding the M5-brane stack
is written as an $S^1_\phi \times S^2_\Omega$ fibration over the $\mu$ interval $[0,1]$.
Clearly, $S^1_\phi$ is identified with the $\phi$ circle in the GM solution \eqref{GM_Toda_frame}, $S^2_\Omega$ is identified with the round $S^2$ in 
\eqref{GM_Toda_frame}, and $\mu$ is identified with $y/N$. 
Furthermore, the connection $v$ in the GM solution is identified
with the \emph{internal} part $A_\Sigma$ of the connection $\cA$ 
on the $N_{SO(2)}$ bundle, $v = - A_\Sigma$, cfr.~\eqref{Dphi_def},
\eqref{NSO2_connection}. By a similar token, the GM 4-form flux
$G_4^{\rm GM}$
is identified with the angular form $E_4$ in \eqref{general_E4}
with all external 4d connections turned off.
More precisely,
\beq \label{GM_is_E4bar}
\frac{G_4^{\rm GM}}{2\pi} = \overline E_4 \qquad \text{in cohomology} \ ,
\eeq 
where the bar over $E_4$ is a reminder that all 4d connections are
switched off.

In order to describe a Riemann surface with punctures,
one has to allow for suitable singular sources in the Toda equation \eqref{Toda_eq}
for $D$. The $\alpha^{\rm th}$ puncture is
described by a source that is  a delta-function localized at a point
$(x_1^\alpha, x_2^\alpha)$ in the $x_1$, $x_2$ directions. 
The profile of the source in the $y$ direction on top of the point 
$(x_1^\alpha, x_2^\alpha)$ encodes the detailed structure of the puncture.
In studying the local geometry near the $\alpha^{\rm th}$ puncture,
it is useful to introduce polar coordinates $r_\Sigma$, $\beta$ via
\beq
x_1 - x_1^\alpha = r_\Sigma \, \cos \beta \  , \qquad
x_2 - x_2^\alpha = r_\Sigma \, \sin \beta \ .
\eeq
In a sufficiently small neighborhood of the puncture,
a rotation of the angle $\beta$ is a symmetry.
Thus, in the study of local puncture geometries
one assumes an additional $U(1)$ rotation symmetry
associated to $\beta$.
Crucially, for a generic
punctured Riemann surface this symmetry does \emph{not} extend
to a \emph{bona fide} isometry of the full solution.

The analysis of solutions to the Toda equation \eqref{Toda_eq}
with additional $U(1)$ symmetry is best performed
by means of the B\"acklund transformation.
The coordinates $(r_\Sigma, y)$ and the function $D = D(r_\Sigma,y)$
are traded for new coordinates $(\rho, \eta)$ and a new function $V = V(\rho,\eta)$
determined implicitly by the relations
\beq \label{backlund_transf}
r_\Sigma^2 \, e^D = \rho^2 \ , \qquad
y = \rho \, \partial_\rho V  \ , \qquad
\log r_\Sigma = \partial_\eta V \ .
\eeq
The source-free Toda equation \eqref{Toda_eq} is mapped
to the source-free, axially symmetric Laplace equation for $V$,
\beq \label{Laplace_eq}
\frac 1 \rho \, \partial_\rho (\rho \, \partial_\rho V) + \partial^2_\eta V = 0 \ .
\eeq
The coordinate $\eta$ parametrizes the axis of cylindrical symmetry,
while $\rho$ is identified with the distance from the axis,
and $\beta$ with the angle around the axis.

The 11d metric and 4-form flux \eqref{GM_Toda_frame}
are written in terms of $\rho$, $\eta$, $V$ as
\begin{align} \label{Backlund_form}
ds^2_{11} & = \kappa^{2/3} \, \bigg[ \frac{\dot V \, \widetilde \Delta}{2 \, V''} \bigg]^{1/3}
\bigg[4 \, ds^2_{AdS_5}
+ \frac{2  \, V'' \, \dot V}{\widetilde \Delta} \, ds^2_{S^2}
+ \frac{2 \, V''}{\dot V} \bigg(
d\rho^2 + d\eta^2 + \frac{2 \, \dot V}{2 \, \dot V - \ddot V} \, \rho^2 \, d\chi^2 \bigg)
\nn \\
& \qquad \qquad \qquad \qquad   + \frac{2 (2 \, \dot V  - \ddot V)}{\dot V \, \widetilde \Delta} \, \bigg(
d\beta - \frac{2 \, \dot V \, \dot V'}{2 \, \dot V  - \ddot V} \, d\chi \bigg)^2 \bigg] \ , \nn \\
G_4^{\rm GM} & = 2 \, \kappa \, {\rm vol}_{S^2} \wedge d \bigg[
- \frac{2 \, \dot V^2 \, V'' }{\widetilde \Delta } \, d\chi  + \bigg(
\eta - \frac{\dot V  \, \dot V'}{ \widetilde \Delta } \bigg) \, d\beta 
\bigg] \ ,
\end{align}
where we used the notation $\dot V = \rho \, \partial_\rho V$, $V' = \partial_\eta V$, and so on, and we
introduced
\beq
\chi = \phi + \beta \ , \qquad
\widetilde \Delta = (2 \, \dot V - \ddot V ) \, V'' + (\dot V')^2 \ .
\eeq
In the presentation \eqref{GM_Toda_frame}, the $S^2$ shrinks
at $y = 0$. After the B\"acklund transformation,
this condition is translated into the boundary condition
$V (\rho, \eta = 0)= 0$.

A puncture is described by a suitable source for the 
Laplace equation \eqref{Laplace_eq}, delta-function localized at $\rho = 0$
and  with  non-trivial 
charge density
profile $\lambda(\eta)$ along the $\eta$ axis. 
The charge density profile $\lambda(\eta)$ is related to $V$ via
\beq
\lambda(\eta) = \lim_{\rho \rightarrow 0} \dot V(\rho, \eta) \ .
\eeq
The analysis of \cite{Gaiotto:2009gz} identifies 
the correct form of $\lambda(\eta)$
corresponding to a
regular puncture. 
Suppose the puncture is labelled by the partition of $N$ determined by
\beq
N = \sum_{a=1}^p \eta_a \, k_a \ , \qquad 
0 < \eta_1 < \eta_2 < \dots < \eta_p    \ , \qquad k_a \ge 1 \ ,
\eeq 
where $\eta_a$ and $k_a$ are integers. The corresponding 
 charge profile $\lambda(\eta)$ is then the continuous piecewise linear
function satisfying
\beq \label{lambda_expr}
\lambda(\eta) = \left\{  \begin{array}{ll}
N_a + \ell_a \, (\eta - \eta_a) & \;\; \eta_{a-1} < \eta < \eta_a \ , \\
N &  \;\; \eta > \eta_p \ ,
\end{array}
\right.
\quad
\ell_a = \sum_{b=a}^p k_b \ , \quad
N_a = \sum_{b=1}^{a-1} \eta_{b} \, k_{b}
+ \eta_a \, \ell_a \ ,
\eeq
where $\eta_0 := 0$.
The explicit solution for $V$ with this source
and satisfying the boundary condition $V(\rho, \eta = 0) = 0$
reads
\beq \label{explicitV}
V(\rho, \eta) = 
 N \, \log \rho + \sum_{a=1}^p \Big[ \cM(\eta_a, k_a)
+ \cM(-\eta_a, -k_a) \Big]  \ ,
\eeq
where
\beq
\cM(\eta_a, k_a) := \frac 12 \, k_a \, \bigg[
(\eta - \eta_a) \, \log \Big( \eta - \eta_a + \sqrt{
\rho^2 + (\eta - \eta_a)^2
}
\Big)  - \sqrt{ \rho^2 + (\eta - \eta_a)^2 }
\bigg] \ .
\eeq
The 11d metric determined by this choice of $V$ 
according to \eqref{Backlund_form} is regular,
up to orbifold singularities of the form $\mathbb R^4/\mathbb Z_{k_a}$
in the four directions $(\rho, \eta, \chi,\beta)$,
located along the $\eta$ axis at $\eta = \eta_a$.
Moreover, the form of $V$ ensures that 
all fluxes of $G_4^{\rm GM}/(2\pi)$
are integrally quantized, if we set $\kappa = (8\pi)^{-1}$
as below \eqref{Nflux_GM}.

The simplest case is $p=1$, corresponding to a partition
of the form $N  = \eta_1 \, k_1$. In this situation
the coordinate transformation relating $(r_\Sigma, y)$
to $(\rho, \eta)$ takes the form
\beq
r_\Sigma = \bigg[ \frac{\eta - \eta_1 + \sqrt{\rho^2 +(\eta - \eta_1)^2 }}{
\eta + \eta_1 + \sqrt{\rho^2 +(\eta + \eta_1)^2 }
} \bigg]^{k_1/2} \ , \qquad
y = \frac{k_1}{2} \bigg[
\sqrt{\rho^2 + (\eta + \eta_1) ^2}
- \sqrt{\rho^2 + (\eta - \eta_1) ^2}
\bigg] \ ,
\eeq
with inverse
\beq \label{inverse_backlund}
\eta = \frac{1 + r_\Sigma^{2/k_1}}{1 - r_\Sigma ^{2/k_1}} \, \frac{y}{k_1} \ , \qquad
\rho = \frac{2 \, r_\Sigma ^{1/k_1} \,  \sqrt{N^2 - y^2}   }{
k_1 \, (1 - r_\Sigma ^{2/k_1})
} \ ,
\eeq
and the function $D$ reads
\beq
e^{D(r_\Sigma,y)} = \frac{4 \, r_\Sigma^{-2+2/k_1} \, (N^2 -y^2)}{k_1^2 \, (1 - r_\Sigma^{2/k_1})^2} \ .
\eeq
If we choose $k_1 = 1$, $\eta_1 = N$, corresponding to the non-puncture,
we recover the expected function $D$ as in \eqref{nonpuncture_D}.

Let us now relate the puncture GM solutions to our inflow setup.
First of all, as already anticipated by our notation,
the B\"acklund transformation \eqref{backlund_transf}
can be regarded as a specific realization of the coordinate change
from the $(r_\Sigma, \mu)$ strip to the $(\rho, \eta)$ quadrant discussed in section 
\ref{sec_nonpuncture_geom} and visualized in figure \ref{plots2}.
Indeed, one verifies that the coordinate transformation
\eqref{inverse_backlund} has the qualitative features
depicted in figure \ref{plots2}. Second of all,
in the metric in \eqref{Backlund_form}
we recognize an $S^1_\beta$ fibration
over the 3d space $(\rho, \eta, \chi)$,
with $\chi = \phi + \beta$ as in the general discussion of section 
\ref{sec_nonpuncture_geom}.
The 3d base space is axially symmetric.
Because of backreaction effects, its metric 
deviates from the flat metric on $\mathbb R^3$,
but one verifies that the 
quantity $2 \dot V /(2 \dot V - \ddot V)$
tends to $1$ as $\rho \rightarrow 0$.
It follows that the $\chi$ circle in the base space
shrinks along the $\eta$ axis in a smooth way. This was the crucial
point in the discussion of section \ref{sec_nonpuncture_geom}.
The connection $L$ for the $S^1_\beta$ fibration,
introduced in \eqref{Dbeta_defin}, is readily read off from  \eqref{Backlund_form},
\beq \label{GM_L}
L = \frac{2 \, \dot V \, \dot V'}{2 \, \dot V  - \ddot V} \ .
\eeq
Using this explicit expression and \eqref{explicitV}
it is easy to verify that $L$ is piecewise constant
along the $\eta$ axis, with jumps located at $\eta = \eta_a$.
The value of $L$ along the interval $(\eta_{a-1}, \eta_a)$
is given by $\ell_a$ as in \eqref{lambda_expr},
which matches exactly with the general relation
\eqref{P_values} derived in section \ref{puncture_geometry}
without reference to the fully backreacted picture.

We can also match the GM 4-form flux in \eqref{Backlund_form}
with the class $E_4$ in the vicinity of the puncture.
It is straightforward to compare \eqref{Backlund_form}
to \eqref{general_E4}, \eqref{new_cE2},  and infer 
\beq
Y + L \, W = \frac{2 \, \dot V^2 \, V''}{\widetilde \Delta} \ , \qquad
W = \eta - \frac{\dot V \, \dot V'}{\widetilde \Delta} \ .
\eeq
Using these explicit expressions, together with \eqref{GM_L},
one can verify that $Y$ and $W$
satisfy the general properties discussed in section \ref{sec_puncture_E4}
without reference to the IR geometry.
In particular, $Y$ is piecewise constant along the $\eta$ axis,
and $Y + L \, W$ is continuous along the $\eta$ axis.
Moreover, one verifies that the quantity $\dot V \, \dot V'/ \widetilde \Delta$
goes to zero at the positions $\eta = \eta_a$. This means that,
in the GM solutions,
\beq
w_a = W(0,  \eta_a) = \eta_a \ . 
\eeq
Of course, the identification of $w_a$ and $\eta_a$ is consistent with the fact that,
in the GM solutions, the locations $\eta_a$ are all integer.
Using $w_a = \eta_a$ we also see a direct match of the expression of $N_a$ 
in \eqref{lambda_expr} with the expression \eqref{Na_expression}
in section \ref{sec_puncture_E4}.
In conclusion, the identification \eqref{GM_is_E4bar}, established earlier
in the absence of punctures, is also valid 
for  puncture geometries. Crucially, even if all 4d connections are turned off,
the class $\overline E_4$ is non-trivial, and encodes the
data that label the puncture.


\section{Proof of Matching with CFT Anomalies}  
\label{sec:matching_proof}

In this appendix we explicitly prove the results  \eqref{eq:result1}-\eqref{eq:result3}.
First, let us evaluate
\begin{align}
(n_v - n_h)^{\rm inflow}(P_\alpha) +(n_v-n_h)^{\text{CFT}}(P_\alpha)  &= \frac 12\,  \sum_{a=1}^p N_a \, k_a  - \frac{1}{2} \sum_{i=1}^{\widetilde{p}} \widetilde{N}_{i} \widetilde{k}_{i} + \frac{1}{2}\ .
\end{align}
	 The quantity $\widetilde{k}_{i}$ is only nonzero at the location of a monopole, which occurs at $i=w_a$. At that location $i=w_a$, $\widetilde{k}_i=k_a$, and $\widetilde{N}_i = N_a$. Then, we can replace
	\ba{
\sum_{i=1}^{\widetilde{p}} \widetilde{N}_{i} \widetilde{k}_{i} 	 = \sum_{a=1}^p N_a k_a\ ,
	}
and the sum simplifies to
	\ba{\bs{
	(n_v - n_h)^{\rm inflow}(P_\alpha) +(n_v-n_h)^{\text{CFT}}(P_\alpha)  &= \frac{1}{2}\ .
	}\label{eq:firstone}
	}
Next, we wish to evaluate
	\begin{align}\bs{
n_v^{\rm inflow}(P_\alpha) + n_v^{\text{CFT}}(P_\alpha) &= 
 \sum_{a=1}^p \bigg[
 \frac 2 3  \, \ell_a^2 \, (w_a^3 - w_{a-1}^3)
+   \ell_a \, (N_a - w_a \, \ell_a) \, (w_a^2 - w_{a-1}^2) \\
&
\qquad - \frac 16 \, N_a \, k_a\bigg]  - \sum_{i=1}^{\widetilde{p}} \left( N^2-\widetilde{N}_{i}^2 \right)  - \frac{1}{2} N^2  + \frac{1}{2}\ . }\label{eq:sub2}
\end{align}
To do this, first note the useful relation
\ba{
\widetilde N_i = N_a + \ell_a (i - w_a) \qquad
\text{for all $i = w_{a-1} , \dots, w_a$} \ .\label{eq:usefull}
}
It follows from \eqref{eq:usefull} and \eqref{Na_expression} that
	\ba{
	\ell_a= \frac{N_a- N_{a-1}}{w_a-w_{a-1}} &= \widetilde{N}_i - \widetilde{N}_{i-1},\qquad i \in [w_{a-1},w_a]\ .
	\label{eq:ells}
	}
	We now re-write the sum over $i$ as a sum over $a$ as
	\ba{
	-\sum_{i=1}^{\widetilde{p}} \left( N^2 - \widetilde{N}_i^2 \right) = - \sum_{a=1}^p \sum_{i=w_{a-1}+1}^{w_a} \left(N^2  - \left[ N_a +\ell_a (i-w_a) \right]^2 \right)\ . \label{eq:toplug}
	}
	Next, we substitute \eqref{eq:toplug} into \eqref{eq:sub2}, pull out a factor of $(w_a-w_{a-1})\ell_a$ where possible in order to make use of the first equality in \eqref{eq:ells}, and  perform the sum over $i$. This gives:
		\ba{\bs{
	\eqref{eq:sub2} = \sum_{a=1}^p & \bigg(\frac{1}{6} \ell_a (N_a-N_{a-1}) + \frac{1}{2} N_a^2 (1+2w_a) - \frac{1}{2} N_{a-1}^2 (1+2w_{a-1}) - \frac{1}{6} k_a N_a\\ &- N^2 (w_a-w_{a-1})\bigg) - \frac{1}{2} N^2 + \frac{1}{2}\ .
	}}
	These sums simplify to
	\ba{
\eqref{eq:sub2} &= \sum_{a=1}^p  \left(	\frac{1}{6} \ell_a (N_a-N_{a-1})  - \frac{1}{6} k_a N_a \right) + \frac{1}{2}\\
&= -\frac{1}{6} \sum_{a=1}^p \left(  \ell_{a+1}N_a - \ell_a N_{a-1}  \right) + \frac{1}{2}\\
& = \frac{1}{2}\ ,
	}
where in the second line we used $k_a=\ell_a-\ell_{a+1}$. Thus we have shown
	\ba{
	n_v^{\rm inflow}(P_\alpha) + n_v^{\text{CFT}}(P_\alpha) &=  \frac{1}{2}\ .\label{eq:secone}
	}
	Together, \eqref{eq:firstone} and \eqref{eq:secone} give the results \eqref{eq:result1} and \eqref{eq:result2} claimed in the main text. The matching of the flavor central charges \eqref{eq:flavorss} and \eqref{eq:cft3} follows from the aforementioned fact that at $i=w_a$, $\widetilde{N}_i=N_a$ and $\widetilde{k}_i = k_a$., and elsewhere $\widetilde{k}_i$ is zero.

\bibliographystyle{./aux/ytphys}
\bibliography{./aux/refs}

\end{document}